\documentclass[epj]{svjour}
\usepackage{amssymb}
\usepackage{amsmath}
\usepackage{array} 
\usepackage{epsfig}
\usepackage{graphics}
\usepackage{colortbl}
\usepackage{hhline}
\usepackage{multirow}
\begin{document}
\title{Visualizations of the Exclusive Central Diffraction}
\author{R.A.~Ryutin\thanks{\emph{e-mail:} Roman.Rioutine@cern.ch}\inst{1}
}                     
%
%
\institute{{\small Institute for High Energy Physics},{\small{\it 142 281}, Protvino, Russia}}
%
%
\abstract{The case of low invariant mass exclusive central 
diffractive production is considered in the general 
theoretical framework. It is 
shown that diffractive patterns (differential cross-sections in variables like
transfer momenta squared, the azimuthal angle between 
final hadrons and their combinations)
can serve as a unique tool to
explore the picture of the $pp$ 
interaction and falsify 
theoretical models. Basic kinematical and dynamical properties of the process
are considered in detail.
As an example, visualizations of diffractive patterns
in the model with three pomerons
for processes $p+p\to p+R+p$ (R is a resonance) 
and $p+p\to p+\pi^{+}\pi^{-}+p$ are 
presented. 
} 
\authorrunning
\titlerunning
\maketitle
%


\section{Introduction}

The central exclusive production process with quasi-diffrac\-ti\-ve\-ly 
scattered initial particles is an important source of information about 
high-energy dynamics of strong interactions both in theory and 
experiment. If we consider only one particle production, this is 
the first "genuinely" inelastic process which not only retains a 
lot of features of elastic scattering but also shows clearly how 
the initial energy is being transformed into the secondary particles.

Theoretical consideration of these processes 
on the basis of Regge theory goes back 
to papers~\cite{EDDEthOLD1}-\cite{EDDEthOLD9}. Experimental 
works were presented in~\cite{EDDEexpOLD1}-\cite{EDDEexpOLD6}. Some 
new interest was related to signals of centrally produced particles like 
Higgs bosons, heavy quarkonia, di-gamma, exotics, 
di-hadrons~\cite{KMR2014}-\cite{EDDEthHH3}. Recent 
data from different experiments
are also available~\cite{EDDEexp1}-\cite{EDDEexp9}. 

In the previous paper~\cite{my2013} the exclusive central 
diffractive production of heavy states was considered in 
detail. In this paper
we present properties of low-mass (central invariant 
masses are less than 3~GeV) exclusive production.

In addition to the general advantages like clear signature with 
two large rapidity gaps (LRG)~\cite{LRGs,LRGs2} and
the possibility to use the ``missing mass method''~\cite{MMM},
there are several specific advantages of the low-mass case. The 
first one is rather large cross-sections. It is 
important, since the schedule for LHC forward physics experiments 
is very limited, and we need also special low luminocity runs
to suppress pile-up events. The second one is the possibility to use 
different diffractive patterns (differential cross-sections on variables like
transfer momenta squared, the azimuthal angle between final hadrons 
and their combinations) as a unique tool to explore the picture 
of the $pp$ interaction and falsify 
theoretical models.

The article is organized as follows. In the first chapter we consider
general kinematical properties and variables of the process. In the second one
we present some model approaches for low-mass exclusive central 
diffraction. In the third part we present visualizations of diffractive patterns
for different processes and kinematical variables and discuss their general 
features. In the conclusions we touch on briefly the future 
experimental possibilities. Appendices are basically devoted to
calculations of amplitudes.

\section{General kinematics and cross-sections}

Let us consider the kinematics of two processes
\begin{eqnarray}
\!\!\!\!\!\!\!\!&&\label{eddeRprod} h_1(p_1)+h_2(p_2)\to h_1(p_1^{\prime})+R(p_R)+h_2(p_2^{\prime}),\\
\!\!\!\!\!\!\!\!&&\label{eddeABprod} h_1(p_1)+h_2(p_2)\to h_1(p_1^{\prime})+\{a(k_a)+b(k_b)\}+h_2(p_2^{\prime}),
\end{eqnarray}
with four-momenta indicated in parentheses. Initial hadrons remain 
intact, $\{a\; b\}$ can be di-boson or di-hadron system and R denotes 
a resonance, ``+'' signs denote large rapidity gaps. Let us call them 
Exclusive Double Diffractive Events (EDDE)
as in our previous 
papers (see~\cite{spin_parity_analyser4} and references 
therein). These processes are also known in the literature 
as Exclusive Double Pomeron Exchange (EDPE), Central Exclusive Diffractive Production (CEDP).

We use the following set of variables:
\begin{eqnarray}
s&=&(p_1+p_2)^2,\; s^{\prime}=(p_1^{\prime}+p_2^{\prime})^2,\; t_{1,2}=(p_{1,2}-p_{1,2}^{\prime})^2,\nonumber\\
s_{1,2}&=&(p_{1,2}^{\prime}+p_R)^2\; \mbox{\rm or}\; (p_{1,2}^{\prime}+k_a+k_b)^2,
\end{eqnarray}

In the light-cone representation $p=\{p_+,p_-; \vec{p}_{\perp}\}$
\begin{eqnarray}
&&\hspace*{-0.3cm}p_1\!=\!\left\{ \sqrt{\frac{\bar{s}}{2}},\frac{m^2}{\sqrt{2\bar{s}}};\; \vec{0}\right\},\;\!\!
\Delta_1\!=\!\left\{ 
\xi_1\sqrt{\frac{\bar{s}}{2}},\frac{-\vec{\Delta}_1^2-\xi_1 m^2}{(1-\xi_1)\sqrt{2\bar{s}}};\; \vec{\Delta}_1
\right\},\;\nonumber\\
&&\hspace*{-0.3cm}p_2\!=\!\left\{ \frac{m^2}{\sqrt{2\bar{s}}},\sqrt{\frac{\bar{s}}{2}};\; \vec{0}\right\},\;\!\!
\Delta_2\!=\!\left\{
\frac{-\vec{\Delta}_2^2-\xi_2 m^2}{(1-\xi_2)\sqrt{2\bar{s}}},\xi_2\sqrt{\frac{\bar{s}}{2}};\; \vec{\Delta}_2
\right\}\nonumber\\
&&\hspace*{-0.3cm}p_{1,2}^{\prime}=p_{1,2}-\Delta_{1,2},\;
p_{1,2}^2=p_{1,2}^{\prime\; 2}=m^2,\nonumber\\
&&\hspace*{-0.3cm} \bar{s}=\frac{s-2m^2}{2}+\frac{s}{2}\sqrt{1-\frac{4m^2}{s}}\simeq s.\label{kin:momenta22}
\end{eqnarray}
Here $\xi_{1,2}$ are fractions of hadrons' longitudinal momenta lost.

Physical region of diffractive events with two large rapidity gaps 
is defined by the following  
kinematical cuts:
\begin{eqnarray}
\label{eq:tlimits}
&&0.01\; GeV^2\le |t_{1,2}|\le\; \sim 1\; GeV^2\;{,} \\
&&\label{eq:xilimits}
\xi_{min}\simeq\frac{M^2}{s \xi_{max}}\le \xi_{1,2}\le \xi_{max}\sim 0.1\;,\\
\label{eq:kappalimits}
&&\left(\sqrt{-t_1}-\sqrt{-t_2}\right)^2\le\kappa\le\left(\sqrt{-t_1}+\sqrt{-t_2}\right)^2\\
&&\kappa=\xi_1\xi_2s-M^2\ll M^2.\nonumber
\end{eqnarray}
$M$ is the invariant mass of the central 
system. We can write the above relations in terms of $y_{1,2}$ (rapidities 
of hadrons), $y$ (rapidity of the central system) and $\eta=(\eta_b-\eta_a)/2$, where $\eta_{a,b}$ are rapidities of particles $a,b$. For instance:
\begin{eqnarray}
&&\label{eq:raplimits}
|y|\le y_0=\ln\left(\frac{\sqrt{s}\xi_{max}}{M}\right),\;
|y_{1,2}|=\frac{1}{2}\ln\frac{(1-\xi_{1,2})^2s}{m^2-t_{1,2}},\nonumber\\
&& |y|\le 6.5,\; |y_{1,2}|\ge 8.75 \mbox{ for } \sqrt{s}=7\;\mathrm{TeV},\nonumber\\
&&|\tanh\eta|\le\sqrt{1-\frac{4m_0^2}{M^2}}.
\end{eqnarray}

Differential cross-sections for the above processes can be represented as
\begin{eqnarray}
&& \frac{d\sigma^{EDDE}_{R}}{d\vec{\Delta}_1^2d\vec{\Delta}_2^2d\phi dy}\simeq\frac{\left| {\cal M}^{EDDE}_R\right|^2}{2^9\pi^4ss^{\prime}},\label{eq:EDDEcsGENR}\\
&& \frac{d\sigma^{EDDE}_{ab}}{d\vec{\Delta}_1^2 d\vec{\Delta}_2^2 d\phi dy dM^2 d\Phi_{ab}}\simeq
\frac{\left| {\cal M}^{EDDE}_{ab}\right|^2}{2^{10}\pi^5 ss^{\prime}},\label{eq:EDDEcsGENab}
\end{eqnarray}
where $\phi$ is the azymuthal angle between outgoing protons, $\Phi_{ab}$ is
the phase space of the dihadron system and ${\cal M}^{EDDE}_{R,\; ab}$ denote 
unitarized amplitudes of the corresponding processes (see ${\cal M}_i^U$ in the Appendix~C).

\section{Double reggeon exchange amplitudes. Approaches.}

\begin{figure}[h!]  
 \includegraphics[width=0.49\textwidth]{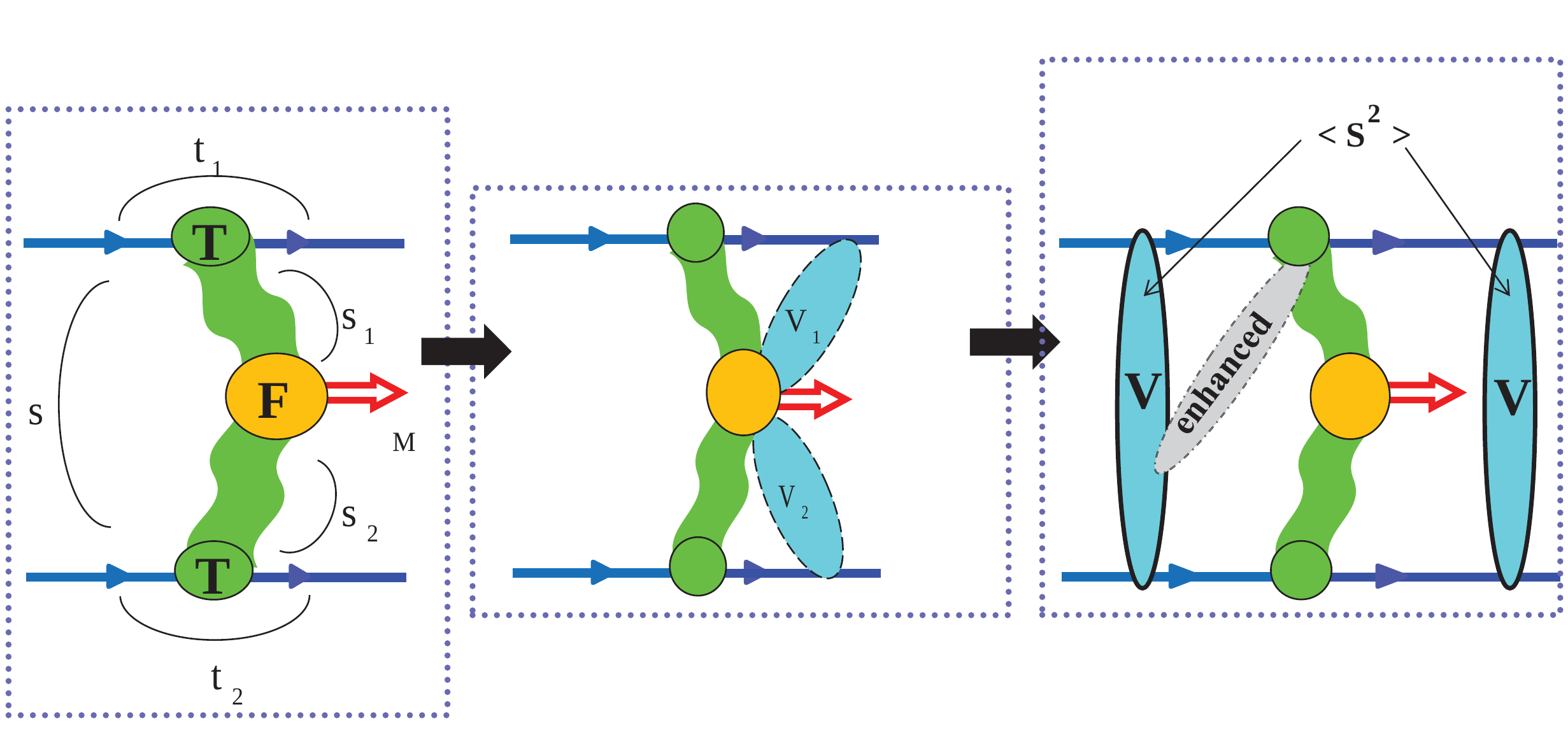} 
  \caption{\label{fig1:lowmalg} Scheme of calculation of the full EDDE amplitude
  in the case of low invariant masses ($M<3$~GeV), i.e. nonperturbative 
  Pomeron-Pomeron fusion.}
\end{figure}

If the central mass produced in EDDE is 
low ($M\sim 1$~GeV, Fig.~\ref{fig1:lowmalg}), it is not
possible to use perturbative representation like in~\cite{my2013} 
for the amplitude of the process, and we have to use 
more general ``nonperturbative'' form. In 
this case we have to obtain somehow the 
Pomeron-Pomeron fusion vertex (see 
Refs.~\cite{my2013,spin_parity_analyser4,spin_parity_analyser2a} 
for details). The scheme of calculations is depicted in 
the Fig.~\ref{fig1:lowmalg}. The first step
is the calculation of the ``bare'' reggeon-reggeon 
amplitude ${\cal M}$, which consists of diffractive form-factors $T$ and
the fusion vertex $F$. If the ``shoulder energies'' $\sqrt{s_{1,2}}$ are 
high enough (say, greater than $100$~GeV), we
also have to take into account rescattering corrections in these 
channels (denoted by $V_{1,2}$). For example, at
$\sqrt{s}=7$~TeV in the kinematical region defined in~(\ref{eq:xilimits}) we 
obtain $~1\;\mathrm{GeV}<\sqrt{s_{1,2}}<2\;\mathrm{TeV}$. Then we should calculate
rescattering corrections in $pp$ channel, which are denoted by $V$. In some 
works~\cite{EDDEth4} they are called ``soft survival probability''. Recently 
it was shown in~\cite{EDDEth4} that enhanced diagrams (additional 
soft interactions) can play significant role.

All the phenomenological models need to obtain va\-lu\-es of their 
parameters to make further predictions. For 
this purpose we can use so called ``standard candle'' processes, i.e. events
which have the same theo\-re\-ti\-cal ingredients for the calculations. For 
low central masses we can use processes:
  \begin{itemize}
    \item $\gamma^*+p\to V+p$ (EVMP), $m_V<3$~GeV~\cite{HERAevmp3}-\cite{HERAevmp4};
    \item $p+p\to p+M+p$, $M=\{q\bar{q}\}$ (light meson) or
 ``glueball''~\cite{EDDEexpOLD1}-\cite{EDDEexpOLD5}, 
  $M=hh$ (dihadron system)~\cite{EDDEexpOLD6}.
  \end{itemize}

From the first principles (covariant reggeization 
approach~\cite{spin_parity_analyser4}) we 
can write the general structure of the vertex for
different cases. For example, for the production
of the low invariant mass system with $J^P$ (spin-parity), when 
$s_i\sqrt{-t_i}\gg \sim 1\;{\mathrm GeV}^3$
and contributions of secondary reggeons are small, we have for
``bare'' amplitudes squared
\begin{eqnarray}
F^{0^{\pm}}_{{\mathbb P}{\mathbb P}}=&&\hspace*{-0.5cm}
\left|
\prod_{i=1,2}
\tilde{T}_0(t_i)
\left(\frac{s_i}{M^2}\right)^{\alpha_i}
\sum_{k=0}^{\infty}
\tilde{\tilde{f}}^k_{0^{\pm}}
\left(
\frac{2\sqrt{t_1t_2}\cos\phi}{M^2}
\right)^k
\right|^2,\nonumber\\
\label{eq:FPOM}
\tilde{T}_0(t_i)&=&
\frac{\alpha^{\prime}_{\mathbb P}}{2} T_0(t_i)
\left( \frac{\sqrt{-t_i}}{m}\right)^{\alpha_i},\\
\tilde{\tilde{f}}^k&=&\tilde{f}^k
\left[
\eta_1\eta_{21}
\Gamma(k-\alpha_1)\Gamma(\alpha_1-
\alpha_2-k)+\right.\nonumber\\
&\phantom{=}&\phantom{\tilde{f}^k[} \left.
\eta_2\eta_{12}
\Gamma(k-\alpha_2)\Gamma(\alpha_2-
\alpha_1-k)
\right],
\end{eqnarray}
\begin{eqnarray}
  \label{eq:cs0plus}\left|{\cal M}^{0^+}\right|^2&\simeq&  
F^{0^+}_{{\mathbb P}{\mathbb P}}, \\
 \label{eq:cs0minus}\left|{\cal M}^{0^-}\right|^2&\simeq&  
F^{0^-}_{{\mathbb P}{\mathbb P}}
\sin^2\phi\;,\\
 \eta_i&=&(-1)^{\sigma_i}+\mathrm{e}^{-\mathrm{i}\pi\alpha_i},\nonumber\\
 \eta_{ij}&=&(-1)^{\sigma_i}(-1)^{\sigma_j}+\mathrm{e}^{-\mathrm{i}\pi(\alpha_i-\alpha_j)},\\
 \alpha_i&=&\alpha_{{\mathbb P}}(t_i),\; \sigma_i=0,
\end{eqnarray}
with 
functions 
defined in the Ap\-pen\-dix~A ($\tilde{f}^k$ are 
nonsingular at $t_i\to 0$, 
$\tilde{T}_0(t)$ is usually represented 
by the exponential $\mathrm{e}^{Bt_i}$ 
or $1/(1-t_i/B)$). Transformation
from integer spins to trajectories was made
like in the Ref.~\cite{Morrow}. 

As one can see from the Appendix~A, 
in the classical Regge scheme
$(-t_i)^{\alpha_i/2}$
is absorbed into the unknown 
residue of the Regge pole. But for 
a fixed integer $J$
this factor always appears 
in the t-channel cosine. In 
Refs.~\cite{spin_parity_analyser3a,spin_parity_analyser3b} results
were obtained from the assumption that the Pomeron acts as a $1^+$ 
conserved or nonconserved current. In particular, it was shown
that the cross-section is proportional to $t_1t_2$, when 
we replace the Pomeron by the conserved
vector current. To remove such zero authors of~\cite{spin_parity_analyser3a}
proposed to use singular functions (nonconserved Pomeron current).

\begin{figure}[h!]  
 \includegraphics[width=0.4\textwidth]{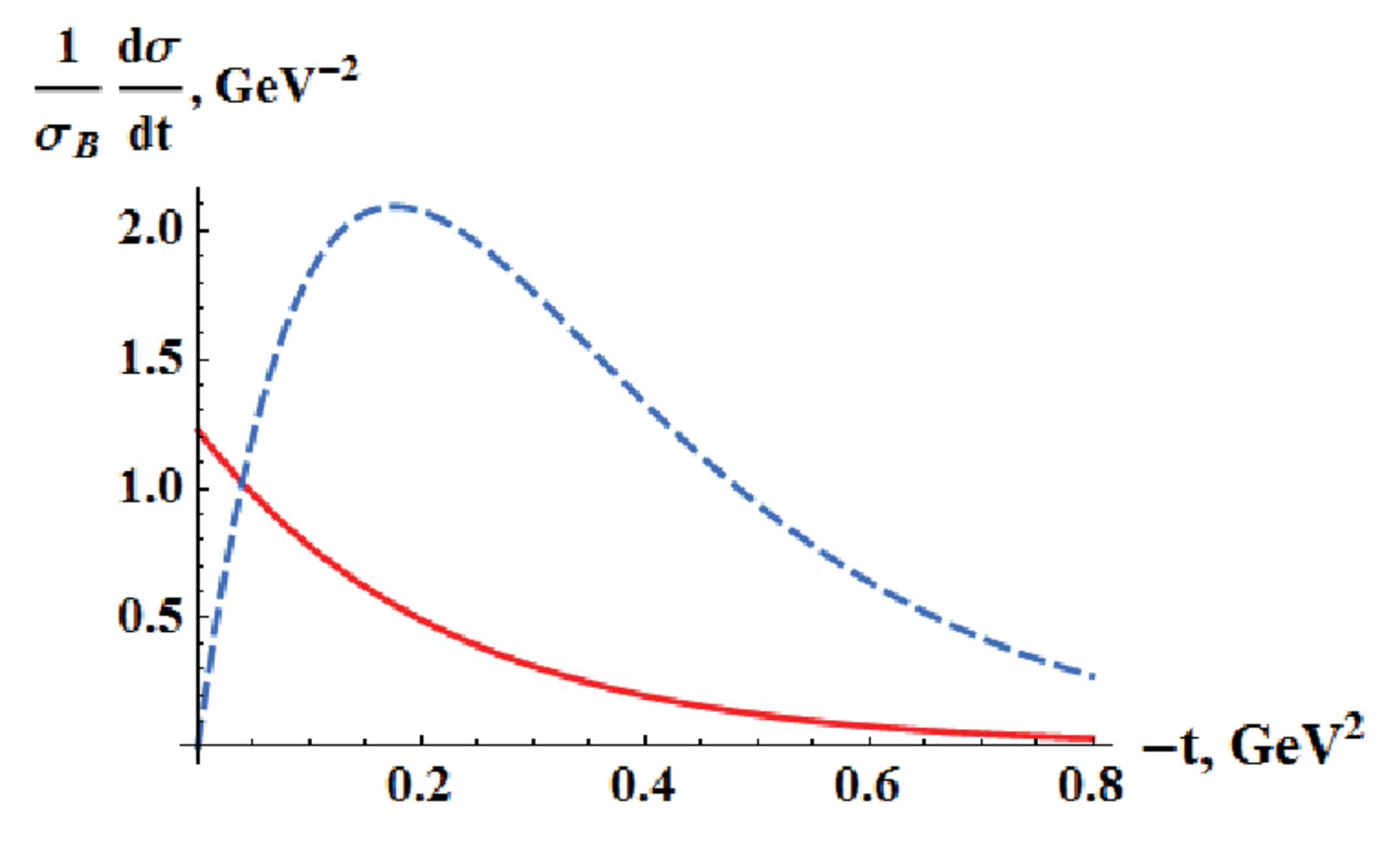} 
  \caption{\label{fig:ZeroRM} The unitarization of the 
  cross-section $|t|{\mathrm e}^{-2B|t|}$ ($B\simeq 2.85\;\mathrm{GeV}^{-2}$, $\sqrt{s}=7$~TeV) corresponding to
  the amplitude~(\ref{vKS}) in the Appendix~C. The dashed curve
  represents the ``bare'' term and the solid one represents
  the unitarized result. $\sigma_B$ is the
  integrated ``bare'' cross-section. The zero at $t=0$ disappears in
  the unitarized cross-section.}
\end{figure}

Strictly speaking, in the real cross-sections rescattering 
corrections at rather high energies can naturally remove
zeroes of a cross-section (see the typical situation in the Fig.~\ref{fig:ZeroRM})
without introducing singular functions.

The general structure
of EDDE amplitudes from the simple Regge behaviour was also 
considered in~\cite{spin_parity_analyser2a,spin_parity_analyser2b} within the
method of helicity amplitudes developed in~\cite{EDDEthOLD5}. 
As was shown in~\cite{spin_parity_analyser4}, experimental data 
are in good agreement with the above predictions.

There were some attempts to obtain the vertex in special 
models. Let us mention first 
the old paper~\cite{Morrow}, where reggeon-reggeon-particle vertex was
exactly calculated in the covariant formalism, and the double 
reggeon amplitude has the form
\begin{eqnarray}
 {\cal M}&\simeq&
\sum_{i\neq j=1}^2 
\frac{\alpha^{\prime}_i\alpha^{\prime}_j}{4}
\left( \frac{s_i}{M^2}\right)^{\alpha_i}
\left( \frac{s_j}{s_0}\right)^{\alpha_j}
\eta_i\eta_{ji}{\cal F}_{ij},\nonumber\\
 {\cal F}_{ij}&=&
\sum\limits_{k=0}^{\infty} \frac{1}{k!}\left( \frac{M_{\perp}^2}{s_0}\right)^k
\Gamma(k-\alpha_i)\Gamma(\alpha_i-\alpha_j-k)=\nonumber\\
\phantom{{\cal F}_{ij}}&\phantom{=}& 
\Gamma(-\alpha_i)\Gamma(\alpha_i-\alpha_j)\!\!
\phantom{F\!\!}_1 F_1(-\alpha_i,1-\alpha_i+\alpha_j; -\frac{M_{\perp}^2}{s_0}),
\nonumber\\
 \label{eq:Morrow}
\alpha_i&=&\alpha^{\prime}_i(0) t_i+\alpha_i(0),
\end{eqnarray}
where $s_0=1\;\mathrm{GeV}^2$ and $\sigma_i$ is the parity of a reggeon. For 
the double Pomeron exchange 
$\alpha_{1,2}=\alpha^{\prime}_{\mathbb P}(0) t_{1,2}+\alpha_{\mathbb P}(0)$. It 
is close to the representation~(\ref{eq:FPOM})  
with exactly calculated couplings. 

The Pomeron-Pomeron fusion based 
on the ``instanton'' or ``glueball''
dynamics was considered in~\cite{instanton1}-\cite{instanton4}. One 
can see also recent papers~\cite{EDDEthH8,lowmassmesonsLHC}
devoted to calculations of the Pomeron-Pomeron fusion vertex in the 
nonperturbative regime.

\section{Diffractive patterns}

Since EDDE is the diffractive process, it retains almost all the features 
of the classical optical diffraction, namely
the diffractive pattern or distribution in the scattering angle. It 
contains the diffractive peak at low angles and
different structures (dips and kinks) at higher angles. Some 
speculations on the meaning of these features can be found
in~\cite{EDDEth6} and further publications. Here we would like to 
point out the following:
\begin{itemize}
\item From the diffractive pattern we extract model independent parameters
of the interaction region such as the $t$-slope which is $R^2/2$, with $R$ 
the transverse radius of the interaction region.
\item We can also estimate the longitudinal size of the 
interaction region~\cite{diff2}:
\begin{eqnarray}
\label{interreg}
&&\Delta x_L>\frac{\sqrt{s}}{2\sqrt{<t^2>-<t>^2}}
\end{eqnarray}
The longitudinal interaction range is somehow "hidden" in the amplitude
but it is this range that is responsible for the "absorption strength". A
rough analogue is the known expression for the radiation absorption
in media which critically depends on the thickness of the absorber.
\item The very presence of dips is the signal of the quantum 
interference of hadronic waves.
\item The depth of dips is determined by the real part of the scattering
amplitude
\end{itemize}

What else could we extract from it? What is the physical meaning 
of the dip position, number of dips or kinks and so on? These
questions stimulate us for future investigations.

\subsection{t-like variables}

In this subsection we present diffractive patterns in t-like variables
for  different physical situations. From the experimental point 
of view it would be more useful to
have distinct structures in 
distributions, since their
position can show the dynamics of 
the interaction and can help to extract parameters with better accuracy.

\begin{figure}[h!]  
 \includegraphics[width=0.49\textwidth]{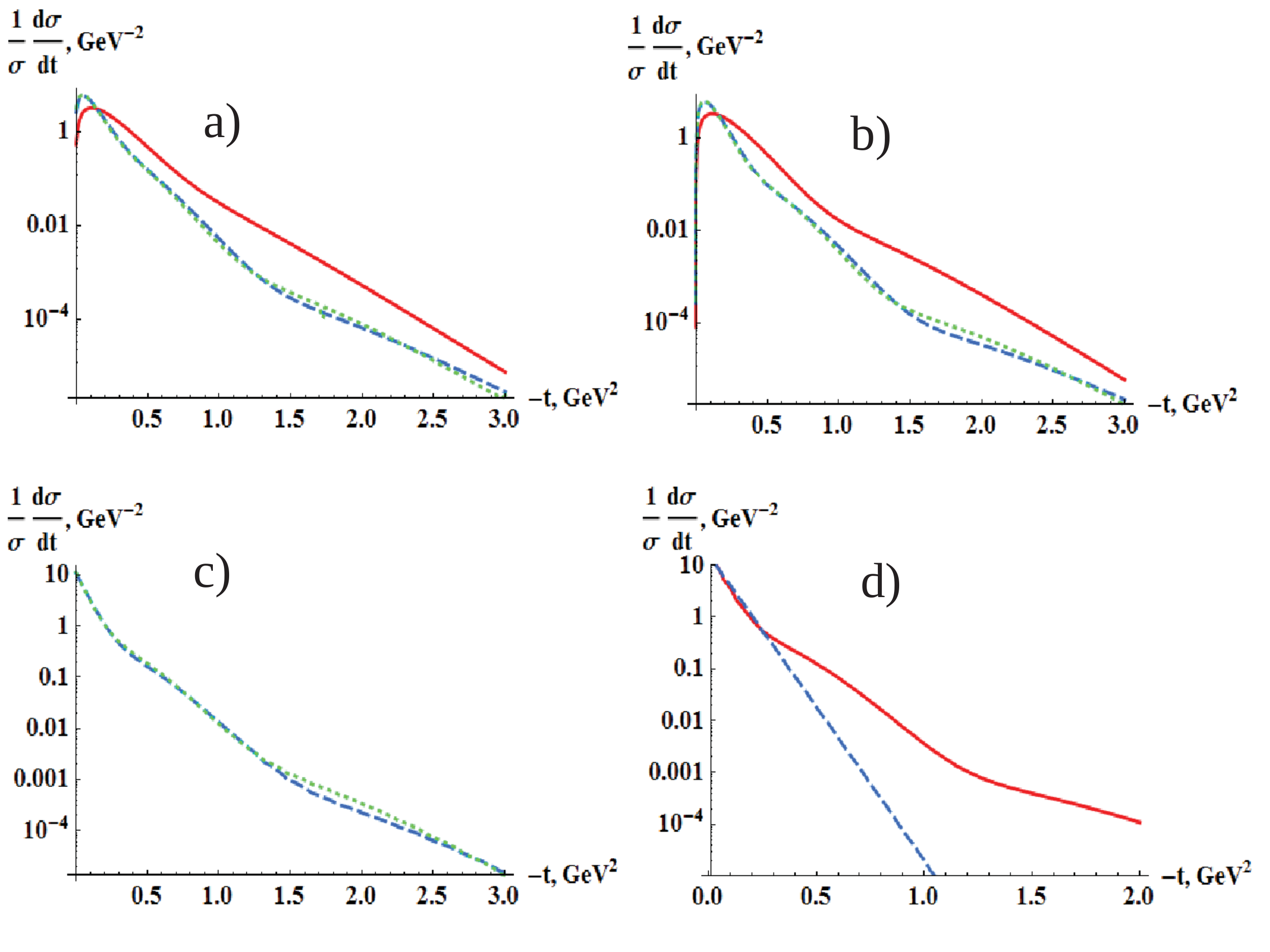} 
  \caption{\label{fig2:tdist} Diffractive t-distributions for different final states (corresponding amplitudes are 
  indicated): a) ``glueball-like''~(\ref{vKS}); b) $\eta^{\prime}$~(\ref{vKV});
  c) $\pi^+\pi^-$~(\ref{eq:MBhh}). Solid curves in a), b) are given 
  for $\sqrt{s}=30$~GeV, dashed and dotted curves in a),b),c) represent
  $\sqrt{s}=7$~TeV and $\sqrt{s}=14$~TeV respectively. Picture d) shows
  the simple ${\mathrm e}^{2Bt}$ cross-section (dashed curve) and the unitarized result (solid curve) at $\sqrt{s}=7$~TeV.}
\end{figure}

In the Fig.~\ref{fig2:tdist} one can see distributions
in t of one of the final protons integrated in other 
variables. Pictures correspond to ``bare'' amplitudes for 
$0^-$~(\ref{vKV}), ``glueball''~(\ref{vKS}) states
and for the pion-pion 
production~(\ref{eq:MBhh}). For 
the simple ${\mathrm e}^{B(t_1+t_2)}$~(\ref{vK0})
amplitude picture~Fig.\ref{fig2:tdist}d) shows 
the signifivance of the
rescattering corrections.

\begin{figure}[h!]  
 \includegraphics[width=0.4\textwidth]{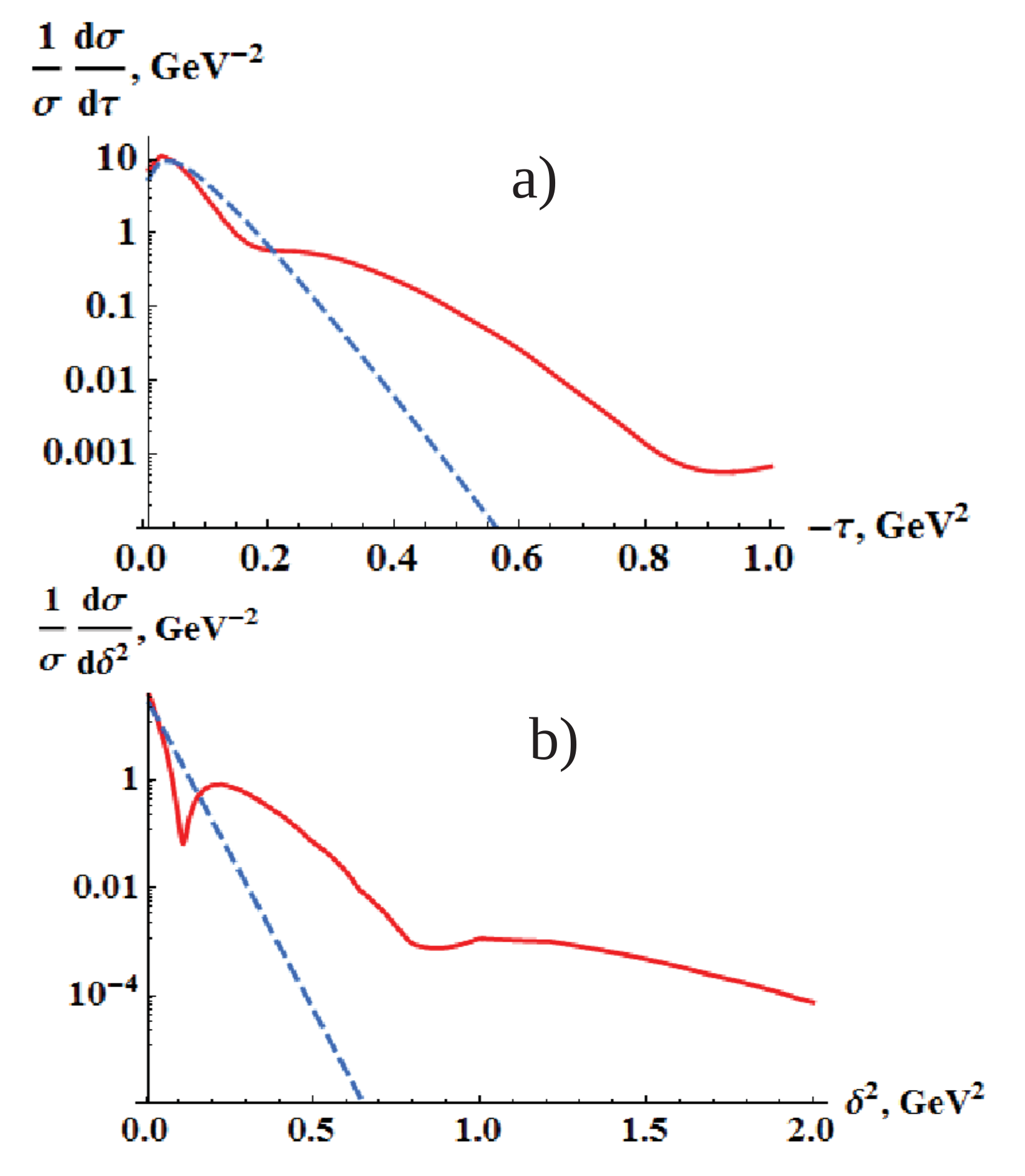} 
  \caption{\label{fig3:othert} Diffractive patterns in different t-like 
  variables: a) $\tau=(t_1+t_2)/2$; b) $\vec{\delta}^2=(\vec{\Delta}_1-\vec{\Delta}_2)^2/4$. Born amplitude 
  (dashed curve) and the unitarized result (solid curve) are shown for 
  $\sqrt{s}=7$~TeV.}
\end{figure}

Let us illustrate how the situation changes, when we use
other variables that seem more natural for the
study of diffractive structures. In the Fig.~\ref{fig3:othert}
we present distributions in $\tau=(t_1+t_2)/2$ and 
$\vec{\delta}^2=(\vec{\Delta}_1-\vec{\Delta}_2)^2/4$
for the case, when the ``bare'' amplitude is the simple 
exponent~(\ref{vK0}) without additional 
structures. For these variables the situation
changes more drastically after taking into account
the unitarization.

\begin{figure}[h!]  
 \includegraphics[width=0.4\textwidth]{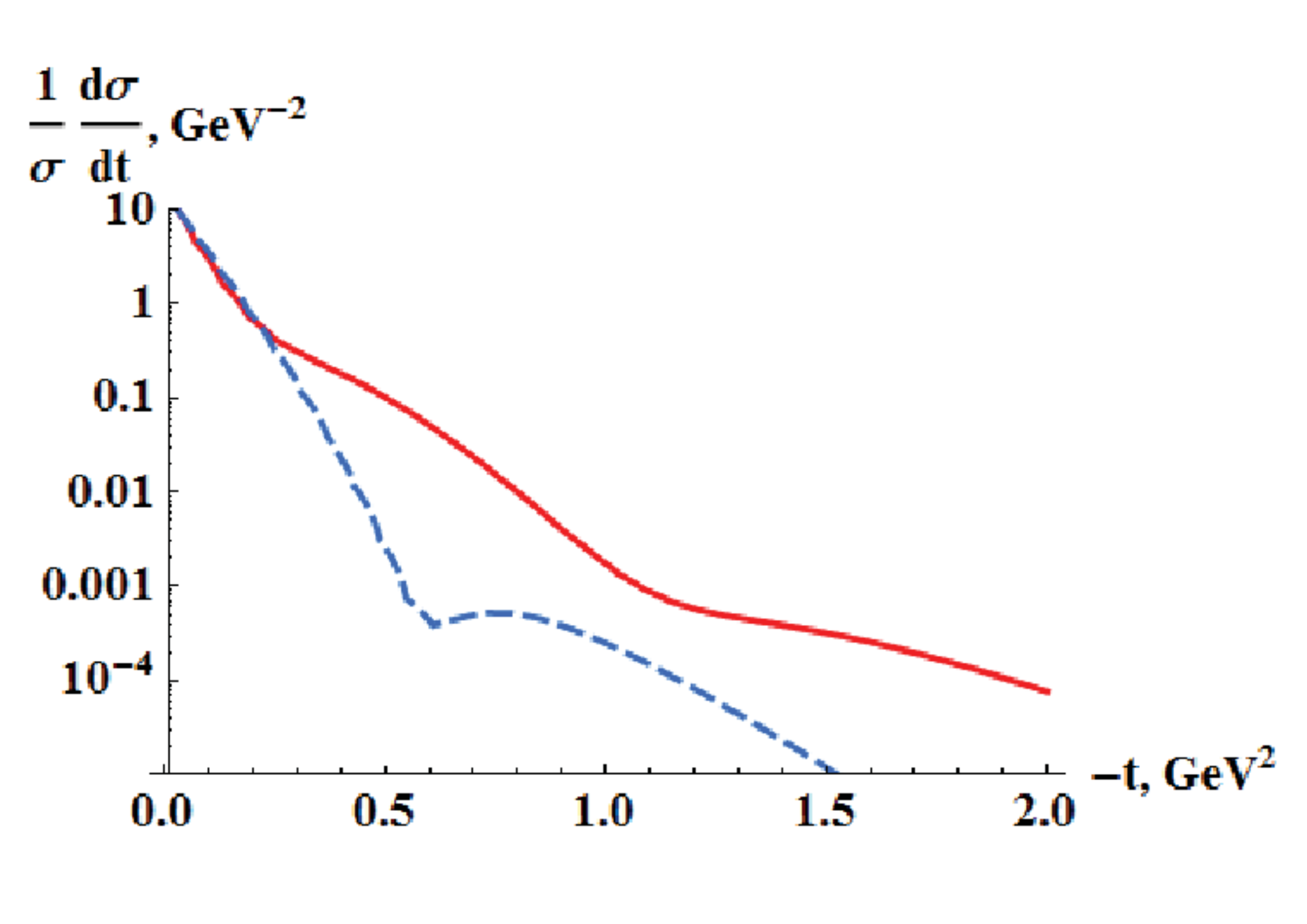} 
  \caption{\label{fig4:dipsmearing} The situation after
  the unitarization (solid curve), when
  the ``bare'' amplitude contains a dip
  structure (dashed curve).}
\end{figure}

On the other hand, as one can see from the 
Fig.~\ref{fig4:dipsmearing}, the effect can 
be the opposite. The ``bare''
amplitude contains the dip at some position, which disappears
in the unitarized distribution, and 
other complicated structure arises. Here we use the toy model
based on the parameters of the third Pomeron from~\cite{3Pom}:
\begin{eqnarray}
&&\label{toy1} {\cal M}\sim \mathrm{e}^{\tilde{B}(t_1+t_2)/2}
\left(\mathrm{e}^{\tilde{B}t_1/2}-\tilde{A}\right)
\left(\mathrm{e}^{\tilde{B}t_2/2}-\tilde{A}\right),\\
&&\label{toy2} \tilde{B}=1.2046+0.5912\left( \ln\left[s (M_{\perp}^2))\right] -\imath\pi\right)/2,
\end{eqnarray}
\begin{eqnarray}
&&\label{toy3} \tilde{A}=49.138 \left(-\imath\sqrt{s}M_{\perp}\right)^{0.0703}/(32\pi\tilde{B}),\\
&&\label{toy4} M_{\perp}^2=M^2-t_1-t_2+2\sqrt{t_1t_2}\cos\phi,\\
&& M=1.5\;\mathrm{GeV},\;\sqrt{s}=7\;\mathrm{TeV}.
\end{eqnarray}


\subsection{Azimuthal correlations}

As was shown earlier in 
references~\cite{spin_parity_analyser2a},\cite{spin_parity_analyser2b}, as 
well as later on in 
references~\cite{spin_parity_analyser4} and~\cite{EDDEthHH1},\cite{EDDEthHH2a}, the 
distribution in the azimuthal angle between final protons can serve
as a powerfull tool to obtain quantum numbers of
centrally produced particles.

\begin{figure}[h!]  
 \includegraphics[width=0.49\textwidth]{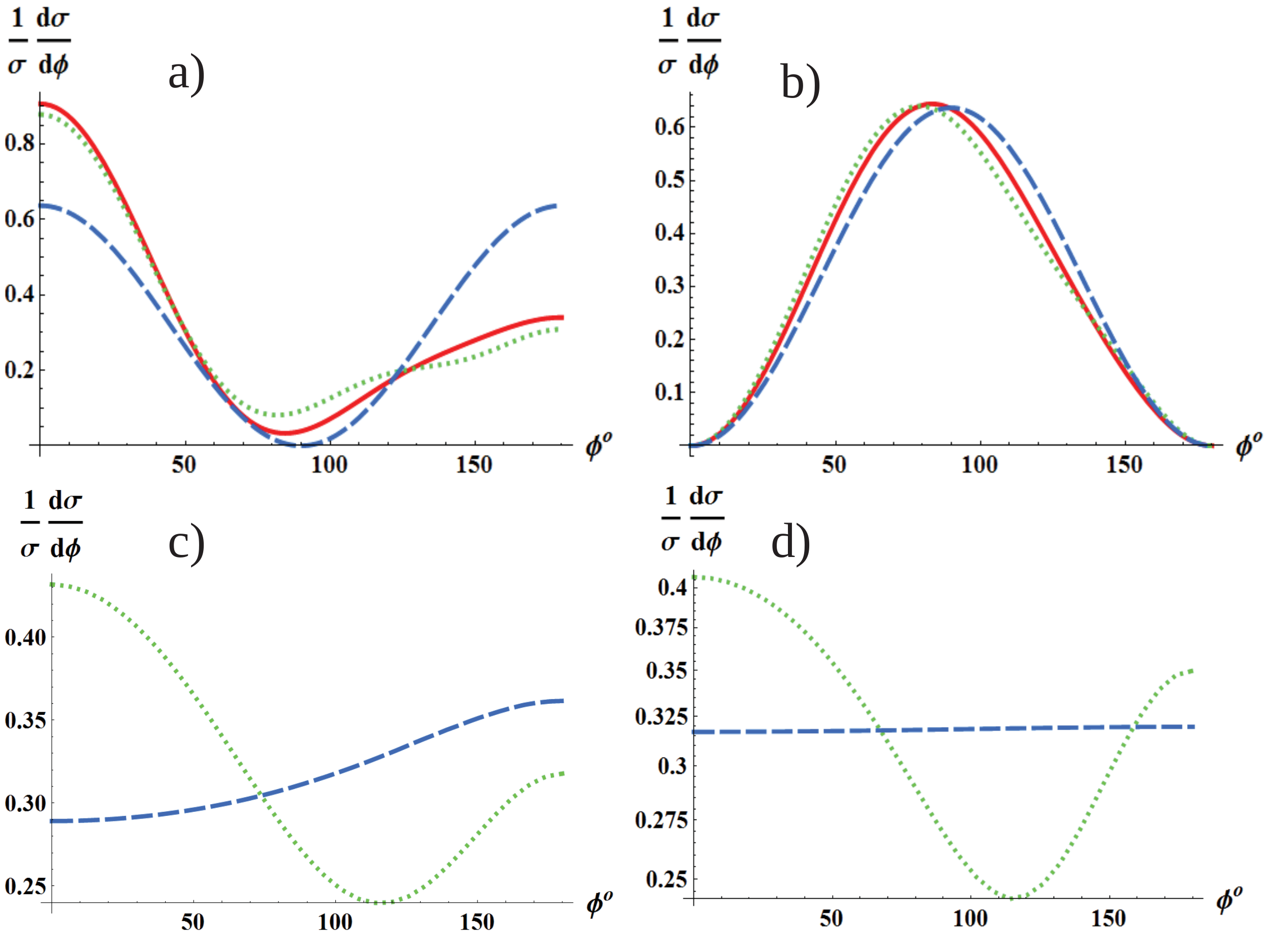} 
  \caption{\label{fig5:phidist} Azymuthal distributions for different
  final states: a) ``glueball-like''~(\ref{vKS}); b) $\eta^{\prime}$~(\ref{vKV});
  c) $\pi^+\pi^-$~(\ref{eq:MBhh}). Solid (red) curves in a), b) are given 
  for $\sqrt{s}=30$~GeV, dotted curves in a),b),c),d) represent unitarized results
  at $\sqrt{s}=7$~TeV. Dashed curves show behaviour of born cross-sections
  at $\sqrt{s}=7$~TeV: 
  a) $\cos^2\phi$, b) $\sin^2\phi$, c) $\pi^+\pi^-$,  d) ``flat''.}
\end{figure}

In the Fig.~\ref{fig5:phidist}a)-c) we present diffractive azimuthal patterns
for $0^-$, $0^+$ (``glueball''), $0^+$ (pion-pion) states. Shapes are 
very different and can be used as a peculiar ``filter''. Furthermore, 
$\phi$-distribution also has strong dependence on the model
that we use for diffractive processes. The unitarization effect for
the ``flat'' distribution
is shown in the Fig.~\ref{fig5:phidist}d). 


\section{Conclusions}
 
 The phenomenon of diffraction is always accompanied by
 specific patterns, partially considered in this paper. We
 have to take it into account when we try to define the
 diffractive process experimentally. Many features of such distributions
 can be very helpful. To continue the 
 paper~\cite{my2013}, here we presented only general aspects of
 the exclusive central production of low invariant mass states, but 
 it is possible to find out
 the same properties in other processes (elastic scattering, single 
 and double diffractive dissociation). For example, we could apply to them
 the procedure of the amplitude construction, which is similar to
 the one stated in the Appendix~A. This hopefully will be
 done in further works.

  As one can see from the above figures, rescattering corrections 
can play significant role and drastically change
the shape of diffractive patterns. We can use this property to 
falsify diffractive models, which are very numerous
``on the market''~\cite{Godizov2012}, with an unprecedented
accuracy.

 Finally, let us mention some possible experimental facilities
 for this task. Since cross-sections of the low mass EDDE
 are rather large ($~10\to 1000\;\mu b$), it is possible
 to use low luminocity runs of the LHC, as was proposed
 in the starting projects~\cite{CMSTOTEM}-\cite{add3}. The recent 
 success of the TOTEM collaboration in t-measurements~\cite{lastTOTEM}
 shows that it is realistic.

\section*{Appendix A}

In this appendix we construct exact reggeon-reggeon fusion amplitudes
for the exclusive production of $0^+$ and $0^-$ states by the
covariant reggeization method proposed in~\cite{spin_parity_analyser4}. For
other states calculations are similar based on formulae for
vertexes from~\cite{spin_parity_analyser4}.

The amplitude $\cal{M}^{J^P}$ (the left picture in the Fig.~\ref{fig1:lowmalg}) 
is composed of vertices $T^{\mu_1\cdots\mu_{J_1}}$, $T^{\nu_1\cdots\nu_{J_2}}$, $F^{\mu_1\dots\mu_{J_1},\;\nu_1\dots\nu_{J_2}}_{\alpha_1\dots\alpha_J}$ and propagators $d(J_i,t)/(m^2(J_i)-t)$ which have the poles at
\begin{equation}
\label{polesJ}
m^2(J_i)-t=0,\; \mbox{i.e.}\; J_i=\alpha_{{\mathbb R}_i}(t)\;,
\end{equation}
after an appropriate analytic continuation of the signatured amplitudes 
in $J_i$. We assume that these poles, where $\alpha_{{\mathbb R}_i}$ are reggeon 
trajectories, give the dominant contribution at high energies after having taken the corresponding residues. Regge-cuts are generated by unitarization.

For vertex
functions $T_{1,2}$ we can obtain the following tensor decomposition:
\begin{eqnarray}
&& T^{(J_i)}\equiv T^{\mu_1\dots\mu_{J_i}}(p_i,\Delta_i)=\nonumber\\
&&\label{Ti} T_0(\Delta_i^2) 
\sum_{n=0}^{\left[\frac{J_i}{2}\right]} 
{\mathbb C}_{J_i}^n \left( P_i^{(J_i-2n)}G_i^{(n)}\right)\;,\\
&&\label{CJn0}{\mathbb C}_{J_i}^n=
\frac{(-1)^n (2(J_i-n))\mbox{!}}{(J_i-n)\mbox{!}n\mbox{!}(J_i-2n)\mbox{!}},
\end{eqnarray}
that satisfies Rarita-Schwinger conditions 
(trans\-verse-sym\-met\-ric-traceless):
\begin{eqnarray}
\label{RaritaShwinger}
&&\Delta_{\mu_i}T^{\mu_1\dots\mu_i\dots\mu_J}=0\label{transverse}\\
&&T^{\mu_1\dots\mu_i\dots\mu_j\dots\mu_J}=T^{\mu_1\dots\mu_j\dots\mu_i\dots\mu_J}\label{symmetric}\\
&&g_{\mu_i\mu_j}T^{\mu_1\dots\mu_i\dots\mu_j\dots\mu_J}=0\label{traceless}
\end{eqnarray}
Tensor structures $\left( P_i^{(J_i-2n)}G_i^{(n)}\right)^{\mu_1\dots\mu_{J_i}}$ satisfy only two 
conditions~(\ref{transverse}),(\ref{symmetric}) (transverse-symmetric) 
and consist of the elements $P_i^{\mu}$ and $G_i^{\mu_1\mu_2}$:
\begin{eqnarray}
&&\label{Pi}P_i^{\;\mu}=
\left( 
p_i^{\mu}-\Delta_i^{\mu}/2
\right)/\sqrt{m^2-\Delta_i^2/4},\\
&&\label{Gi} G_i^{\mu_1\mu_2}=g^{\mu_1\mu_2}-
\frac{\Delta_i^{\mu_1}\Delta_i^{\mu_2}}{\Delta_i^2}\;.
\end{eqnarray}
\begin{eqnarray}
&&\left( P_i^{(J_i-2n)}G_i^{(n)}\right)=\nonumber\\
&&\label{PGsym} \frac{P_i^{\; (\mu_1}\!\!\cdot_{\dots}\!\!\cdot P_i^{\; \mu_{J_i-2n}}
G_i^{\mu_{J_i-2n+1}\mu_{J_i-2n+2}}\!\!\cdot_{\dots}\!\!\cdot G_i^{\mu_{J_i-1}\mu_{J_i})}}{N_{J_i}^n},\\
&& \label{coefPGsym}N_{J_i}^n=\frac{J_i\mbox{!}}{2^n n\mbox{!}(J_i-2n)\mbox{!}}.
\end{eqnarray}
Coefficients ${\mathbb C}_{J_i}^n$ in~(\ref{Ti}) can be obtained from 
the condition~(\ref{traceless}), which
leads to the recurrent set of equations. For each transverse-symmetric
structure we have
\begin{eqnarray}
&&g_{\mu_1\mu_2}\left( P_i^{(J_i-2n)}G_i^{(n)}\right)^{\mu_1\dots\mu_{J_i}}=
\nonumber\\
&& \frac{(J_i-2n)(J_i-2n-1)}{J_i(J_i-1)}\times
\left( P_i^{(J_i-2n-2)}G_i^{(n)}\right)+\nonumber\\
&& \frac{2n(2(J_i-2)N_{J_i-3}^{n-1}+(J_i-2)(J_i-3)N_{J_i-4}^{n-2}+3N_{J_i-2}^{n-1})}{J_i(J_i-1)N_{J_i-2}^{n-1}}\times\nonumber\\
&& \label{traceTmu}\left( P_i^{(J_i-2n)}G_i^{(n-1)}\right)=0,
\end{eqnarray}
where the first term corresponds to the tensor contraction 
\begin{equation}
P_i^{\mu_1}P_i^{\mu_2}g_{\mu_1\mu_2}=1,
\end{equation}
and three items 
in the second term correspond to 
\begin{eqnarray}
P_i^{\mu_1} G_i^{\mu_2\mu_k}g_{\mu_1\mu_2}&=&P_i^{\mu_k},\\
G_i^{\mu_1\mu_k} G_i^{\mu_2\mu_l}g_{\mu_1\mu_2}&=&G_i^{\mu_k\mu_l},\\
G_i^{\mu_1\mu_2}g_{\mu_1\mu_2}&=&3.
\end{eqnarray}
Finally, we have
\begin{eqnarray}
&& \sum_{n=0}^{\left[\frac{J_i}{2}\right]} {\mathbb C}_{J_i}^n\times
\left[
(J_i-2n)(J_i-2n-1)\times
\left( P_i^{(J_i-2n-2)}G_i^{(n)}\right)+
\right.\nonumber\\
&&\label{traceTmu2}\left.
2n (2J_i-2n+1)\times
\left( P_i^{(J_i-2n)}G_i^{(n-1)}\right)
\right]=\nonumber\\
&& \sum_{n=1}^{\left[\frac{J_i}{2}\right]}
\left[
{\mathbb C}_{J_i}^{n-1}(J_i-2n+2)(J_i-2n+1)+
\right.\nonumber\\
&&\label{traceTmu3}\left.
{\mathbb C}_{J_i}^{n}2n(2J_i-2n+1)
\right]\times \left( P_i^{(J_i-2n)}G_i^{(n-1)}\right)=0
\end{eqnarray}
and
\begin{eqnarray}
&&{\mathbb C}_{J_i}^n={\mathbb C}_{J_i}^{n-1}
\frac{(-1)(J_i-2n+2)(J_i-2n+1)}{2n(2J_i-2n+1)}=\nonumber\\
&&\label{CJn} \frac{(-1)^n (2(J_i-n))\mbox{!}}{(J_i-n)\mbox{!}n\mbox{!}(J_i-2n)\mbox{!}}\left[ {\mathbb C}_{J_i}^0\frac{(J_i\mbox{!})^2}{(2J_i)\mbox{!}}\right],
\end{eqnarray}
which is equal to~(\ref{CJn0}), if we set the expression in square
brackets to unity.

Now let us obtain the general expression for the vertex
$F^{(J_1),(J_2)}_{(J)}\equiv
F^{\mu_1\dots\mu_{J_1},\;\nu_1\dots\nu_{J_2}}_{\alpha_1\dots\alpha_J}$, when 
$J=0$. Since this tensor has to satisfy~(\ref{transverse})-(\ref{traceless}) in
each group of indexes, it should be represented as
\begin{eqnarray}
&& F^{(J_1),(J_2)}_{0^+}=\sum_{k=0}^{\min [J_1,J_2]}
\sum_{n_i=0}^{\left[\frac{J_i-k}{2}\right]} 
{\mathbb C}_{J_1,J_2}^{k,n_1,n_2}\times\nonumber\\
&& \label{FJ1J20plus0}\left( 
D_1^{(J_1-k-2n_1)}G_1^{(n_1)}G_{12}^{(k)}G_2^{(n_2)}D_2^{(J_2-k-2n_2)} 
\right).
\end{eqnarray}
Here the transverse-symmetric structure in parentheses 
contains two groups of indexes: 
$\{\mu\}\equiv\mu_1\dots\mu_{J_1}$ and $\{\nu\}\equiv\nu_1\dots\nu_{J_2}$
and consists of the following elements:
\begin{eqnarray}
&&\label{Di} D_{1,2}^{\rho}=\left( \Delta_{1,2}^{\rho}+
\frac{d_{1,2}^2}{(\Delta_1\Delta_2)}\Delta_{2,1}^{\rho}\right)/\left(d_{1,2} K_{12}\right),\\
&&\label{diK12} d_{1,2}=\sqrt{-t_{1,2}},\; 
K_{12}=\sqrt{1-\frac{d_1^2d_2^2}{(\Delta_1\Delta_2)^2}},\\
&&\label{G12} G_{12}^{\mu\nu}=g^{\mu\nu}-
\frac{\Delta_2^{\mu}\Delta_1^{\nu}}{(\Delta_1\Delta_2)},
\end{eqnarray}
and $G_i$ is defined in~(\ref{Gi}). The number of different 
terms in each structure is
\begin{equation}
N_{J_1J_2}^{k,n_1,n_2}=\frac{N_{J_1}^{n_1}N_{J_2}^{n_2}}{k\mbox{!}}.
\end{equation}
For $0^-$ state we have to add
the anti-symmetric element
\begin{equation}
F_A^{\mu\nu}=\epsilon^{\mu\nu\rho\sigma}\Delta_{1,\rho}\Delta_{2,\sigma}/(d_1d_2),
\end{equation}
and the vertex looks as follows:
\begin{eqnarray}
\hspace*{-0.6cm}&& F^{(J_1),(J_2)}_{0^-}=\sum_{k=0}^{\min [J_1,J_2]-1}
\sum_{n_i=0}^{\left[\frac{J_i-1-k}{2}\right]} 
{\mathbb C}_{J_1,J_2}^{k,n_1,n_2}\times\nonumber\\
\hspace*{-0.6cm}&& \label{FJ1J20minus0}\left( 
F_A D_1^{(J_1-1-k-2n_1)}G_1^{(n_1)}G_{12}^{(k)}G_2^{(n_2)}D_2^{(J_2-1-k-2n_2)} 
\right).
\end{eqnarray}
For further calculations let us define additional quantities and 
functions (approximate values are given for 
$d_{1,2}\ll m\le M\ll \sqrt{s_{1,2}}$):
\begin{eqnarray}
\label{vXi}
X_{1,2}&=&
\frac{(P_{1,2}\Delta_{2,1})\; d_{1,2}}{(\Delta_1\Delta_2)\; Q_{1,2}}
\simeq \frac{s_{1,2}\;d_{1,2}}{M^2\; m},\\
\label{vQi} Q_{1,2}&=&\sqrt{1+d_{1,2}^2/(\Delta_1\Delta_2)},\\
\label{vZ12} Z_{12}&=&
\frac{(P_1P_2)\; (\Delta_1\Delta_2)}{(P_1\Delta_2)\; (P_2\Delta_1)}\simeq
1-\frac{2\vec{\Delta}_1\vec{\Delta}_2}{M^2},\\
\label{fk} {\mathbb C}_{J_1J_2}^{k,0,0}&=&
\left(\frac{d_1d_2}{(\Delta_1\Delta_2)}\right)^k f_{J_1J_2}^k, 
\end{eqnarray}
where $f_{J_1J_2}^k$ are nonsingular at $t_i\to 0$ 
functions of $t_1$, $t_2$ and $M^2$.

We can construct $F^{(J_1)(J_2)}_{0^{\pm}}$ vertexes as we did
for $T^{(J_i)}$ in~(\ref{traceTmu})-(\ref{CJn}), taking the trace
in each group of indexes and obtaining recurrent equations
for ${\mathbb C}_{J_1J_2}^{k,n_1,n_2}$. It will be done in further
works. Here we note, that in the contraction
\begin{equation}
\label{TFTcontr1}
V_{J_1J_2,\; 0^{\pm}}=\frac{T^{(J_1)}_{\{\mu\}}}{T_0(t_1)}\otimes 
F^{(J_1),\;\{\mu\}\;(J_2),\;\{\nu\}}_{0^{\pm}}\otimes 
\frac{T^{(J_2)}_{\{\nu\}}}{T_0(t_2)}
\end{equation}
$F$-vertexes can be replaced by
\begin{eqnarray}
&&\hspace*{-0.7cm} F^{(J_1),(J_2)}_{0^+}\to 
\left( \frac{d_1}{(\Delta_1\Delta_2)K_{12}}\right)^{J_1}
\left( \frac{d_2}{(\Delta_1\Delta_2)K_{12}}\right)^{J_2}\times\nonumber\\
&&\hspace*{-0.7cm}\phantom{F^{(J_1),(J_2)}_{0^+}\to}
\sum_{k=0}^{\min [J_1,J_2]}
f_{J_1,J_2}^{k}
\left( (\Delta_1\Delta_2)K_{12}^2\right)^k
\times\nonumber\\
&&\hspace*{-0.7cm} \label{FJ1J20plus1}
\Delta_2^{\mu_{k+1}}\!\cdot_{\dots}\!\!\cdot\Delta_2^{\mu_{J_1}}
G_{12}^{\mu_1\nu_1}\!\cdot_{\dots}\!\!\cdot G_{12}^{\mu_k\nu_k}
\Delta_1^{\nu_{k+1}}\!\cdot_{\dots}\!\!\cdot\Delta_1^{\nu_{J_2}},\\
&&\hspace*{-0.7cm} F^{(J_1),(J_2)}_{0^-}\to 
\left( \frac{d_1}{(\Delta_1\Delta_2)K_{12}}\right)^{J_1}
\left( \frac{d_2}{(\Delta_1\Delta_2)K_{12}}\right)^{J_2}
\times\nonumber\\
&&\hspace*{-0.7cm}\phantom{F^{(J_1),(J_2)}_{0^-}\to}
\sum_{k=0}^{\min [J_1-1,J_2-1]}
f_{J_1,J_2}^{k}
\left( (\Delta_1\Delta_2)K_{12}^2\right)^k
\times\nonumber\\
&&\hspace*{-0.7cm} \label{FJ1J20minus1} 
F_A^{\mu_1\nu_1}\Delta_2^{\mu_{k+2}}\!\cdot_{\dots}\!\!\cdot\Delta_2^{\mu_{J_1}}
G_{12}^{\mu_2\nu_2}\!\cdot_{\dots}\!\!\cdot G_{12}^{\mu_{k+1}\nu_{k+1}}
\Delta_1^{\nu_{k+2}}\!\cdot_{\dots}\!\!\cdot\Delta_1^{\nu_{J_2}},
\end{eqnarray}
due to transverse-symmetric-traceless properties of $T$ 
struc\-tu\-res. It is
possible to show that in the exact $F$-vertexes 
coefficients ${\mathbb C}_{J_1J_2}^{k,n_1,n_2}$, $n_i> 0$ can be
expressed in terms of $f_{J_1J_2}^k$ only, i. e. we can
obtain the exact formulae for~(\ref{TFTcontr1}) by the use 
of simplified 
expansions~(\ref{FJ1J20plus1}),(\ref{FJ1J20minus1}), which is done below.

Let us calculate leading terms in the expansions of contracted
vertexes
\begin{eqnarray}
&&\label{TFTcontr0plus}
V_{J_1J_2,\; 0^+}=\sum_{k=0}^{\min[J_1,J_2]} V^k_{J_1J_2,\; 0^+},\\
&&\label{TFTcontr0minus}
V_{J_1J_2,\; 0^-}=\sum_{k=0}^{\min[J_1-1,J_2-1]} V^k_{J_1J_2,\; 0^-}.
\end{eqnarray}
It is rather easy to show that
\begin{eqnarray}
&& V^0_{J_1J_2,\; 0^+}=f_{J_1J_2}^0
\prod_{i=1}^2
\left(\frac{2Q_i}{K_{12}}\right)^{J_i}{\cal P}_{J_i}(X_i)\simeq\nonumber\\
&& \label{V0plus0}
\tilde{f}^0_{J_1J_2} X_1^{J_1}X_2^{J_2}\mbox{ for } X_i\gg 1,\; t_i\ll m^2,
\end{eqnarray}
where ${\cal P}_J(X)$ are Legendre polynomials and numerical
factors are absorbed into 
$\tilde{f}^0_{J_1J_2}$. For the next term we can apply 
the following trick
\begin{eqnarray}
&& \left( P_1^{(J_1-2n_1)}G_1^{(n_1)}\right)_{\{\mu\}}=\nonumber\\
&& P_1^{\mu_1}\frac{J_1-2n_1}{J_1}
\left( P_1^{(J_1-2n_1-1)}G_1^{(n_1)}\right)^{\{\mu\}\neq\mu_1}+\nonumber\\
&& \label{trickV1}\sum_{i=2}^{J_1}G_1^{\mu_1\mu_i}\frac{2n_1}{J_1}
\left( P_1^{(J_1-2n_1)}G_1^{(n_1-1)}\right)^{\{\mu\}\neq\mu_1,\mu_i}
\end{eqnarray}
and the same for the second structure. Effectively in the
contaction the following dimensionless factor has to be added
\begin{eqnarray}
&&\left[ 
\frac{(J_1-2n_1)\sqrt{(\Delta_1\Delta_2)}}{J_1(P_1\Delta_2)}P_1^{\mu_1}+
\frac{(2n_1)d_1 K_{12}}{J_1\sqrt{(\Delta_1\Delta_2)}Q_1^2}D_1^{\mu_1}
\right]\times\nonumber\\
&& \label{effexp1}\left[\phantom{\frac{J_1-2n_1}{J_1(P_1\Delta_2)}P_1^{\mu_1}}
\hspace*{-2.2cm}1\to 2,\; \mu\to\nu
\right].
\end{eqnarray}
Then we have to contract
these structures with $G_{12}^{\mu_1\nu_1}$ and $F_A^{\mu_1\nu_1}$ to
calculate $V^1_{J_1J_2,\; 0^+}$ and $V^0_{J_1J_2,\; 0^-}$
respectively. The final result can be represented as
\begin{eqnarray}
&& V^1_{J_1J_2,\; 0^+}=
f_{J_1J_2}^1
\left(\frac{2Q_1}{K_{12}}\right)^{J_1}
\left(\frac{2Q_2}{K_{12}}\right)^{J_2}
\frac{K_{12}^2}{J_1J_2}\times\nonumber\\
&&\left\{
\left[ Z_{12}-1 \right]{\cal P}^{\prime}_{J_1}{\cal P}^{\prime}_{J_2}+
\frac{d_1^2d_2^2}{(\Delta_1\Delta_2)^2}
\left[
\frac{K_{12}^2}{Q_1^2Q_2^2}
{\cal P}^{\prime\prime}_{J_1}{\cal P}^{\prime\prime}_{J_2}+
\right.
\right.
\nonumber\\
&& \left.
\left.\phantom{\frac{K_{1}^2}{Q_1^2}}
\frac{1}{Q_2^2}{\cal P}^{\prime}_{J_1}{\cal P}^{\prime\prime}_{J_2}+
\frac{1}{Q_1^2}{\cal P}^{\prime\prime}_{J_1}{\cal P}^{\prime}_{J_2}
\right]
\right\}\simeq\nonumber\\
&&\label{V0plus1}\phantom{V^1_{J_1J_2,\; 0^+}=}
\tilde{f}^1_{J_1J_2}X_1^{J_1}X_2^{J_2}
\frac{2\vec{\Delta}_1\vec{\Delta}_2}{M^2},
\end{eqnarray}

\begin{eqnarray}
&& V^0_{J_1J_2,\; 0^-}=f_{J_1J_2}^0
\left\{ \frac{\sqrt{1-\frac{4m^2}{s}}}{1-
\frac{s_1+s_2}{2s}+\frac{M^2-4m^2-d_1^2-d_2^2}{4s}}\right\}
\times\nonumber\\
&& 
\left(\frac{2Q_1}{K_{12}}\right)^{J_1}
\left(\frac{2Q_2}{K_{12}}\right)^{J_2}
\frac{Z_{12}}{J_1J_2}
{\cal P}^{\prime}_{J_1}{\cal P}^{\prime}_{J_2}
\frac{\left[ \vec{\Delta}_1\times\vec{\Delta}_2\right]}{d_1d_2}\simeq\nonumber\\
&& \label{V0minus0}\phantom{V^0_{J_1J_2,\; 0^-}=}
\tilde{f}^0_{J_1J_2}X_1^{J_1}X_2^{J_2}
\frac{\left[ \vec{\Delta}_1\times\vec{\Delta}_2\right]}{d_1d_2},
\end{eqnarray}
where the term in braces is close to unity and
\begin{eqnarray}
&&
{\cal P}^{\prime}_{J}=
X\frac{\partial}{\partial X}{\cal P}_{J}(X)=\nonumber\\
&&\label{PJpr} 
\frac{J\;X}{X^2-1}\left( 
X\;{\cal P}_{J}(X)-
{\cal P}_{J-1}(X)
\right),\\
&& 
{\cal P}^{\prime\prime}_{J}=
-\frac{\partial}{\partial X}{\cal P}_{J-1}(X)=\nonumber\\
&&\label{PJprpr}
-\frac{J}{X^2-1}\left( 
{\cal P}_{J}(X)-X\;
{\cal P}_{J-1}(X)
\right).
\end{eqnarray}
For $d_{1,2}\ll m\le M\ll \sqrt{s_{1,2}}\le\sqrt{s}$ and $X_i\gg 1$
we can write the expressions for leading terms of amplitudes
\begin{eqnarray}
&& 
{\cal M}^{0^+}\simeq 
\sum_{J_1,J_2}
\prod_{i=1,2} \left[
T_0(t_i)
X_i^{J_i}
\right]\times\nonumber\\
&& \label{cM0plus}
\sum_{k=0}^{\min(J_1,J_2)}
\tilde{f}^k_{J_1J_2}\left(
\frac{2\sqrt{t_1t_2}\cos\phi}{M^2}
\right)^k,\\
&& 
{\cal M}^{0^-}\simeq 
\sum_{J_1,J_2}
\prod_{i=1,2} \left[
T_0(t_i)
X_i^{J_i}
\right]\times\nonumber\\
&& \label{cM0minus}
\sum_{k=0}^{\min(J_1,J_2)-1}
\tilde{f}^k_{J_1J_2}\left(
\frac{2\sqrt{t_1t_2}\cos\phi}{M^2}
\right)^k
\sin\phi.
\end{eqnarray}
Then we have to continue analytically the above expressions 
to complex $J_{1,2}$ 
planes. It can be done like in the Ref.~\cite{Morrow}, using
the reggeization prescription
\begin{equation}
\label{reggeizationpr}
\sum_J\frac{F^J}{(t-m^2)}\to 
\frac{\alpha_{{\mathbb R}}^{\prime}}{2}\eta_{{\mathbb R}}(t)\Gamma(-\alpha_{{\mathbb R}}(t))F^{\alpha_{{\mathbb R}}(t)}.
\end{equation}

To check that the above approach coincides with the usual Regge one, let us
calculate the amplitude of the elastic scattering of two particles with equal
masses $m$. For the exchange of the 
meson with spin $J$ it is equal to the contraction
\begin{eqnarray}
&&{\cal M}^{el}(s,t)=T^{(J)}_{\{\mu\}}(p_1,\Delta)\otimes 
T^{(J)}_{\{\mu\}}(p_2,-\Delta)=\nonumber\\
&& \label{elasticJ}T_0(t)^2 2^J {\mathbb C}_J^0 
{\cal P}_J\left( 
\frac{s-2m^2+t/2}{2m^2-t/2}
\right)\sim \left(\frac{s}{m^2}\right)^J,
\end{eqnarray}
which leads to the basic reggeon exchange formula after appropriate
analytical continuation to the complex $J$ plane. 

More complicated
situation occurs in the case of unequal masses. For example, let us
consider the process $p+p\to p+X$, where $m_p=m$, $m_X=M\gg m$. For the exchange 
of the meson with spin $J$ we have
\begin{eqnarray}
&&{\cal M}(s,t)=T^{(J)}_{\{\mu\}}(p_1,\Delta)\otimes 
T^{(J)}_{\{\mu\}}(p_2,-\Delta)=\nonumber\\
&& T_{01}(t)T_{02}(t) 2^J {\mathbb C}_J^0 
{\cal P}_J\left( 
\frac{(2s-3m^2-M^2+t)\sqrt{-t}}{\sqrt{4m^2-t}\;\lambda^{1/2}(t,m^2,M^2)}
\right) \nonumber\\
&&\phantom{{\cal M}(s,t)}\label{UneqMass} \sim \left(\frac{s\sqrt{-t}}{M^2m}\right)^J.
\end{eqnarray}
Here the argument of the Legendre function is the t-channel cosine $z_t=\cos\theta_t$, and 
$$
\lambda(x,y,z)=x^2+y^2+z^2-2xy-2xz-2yz.
$$ 
Factor $\sqrt{-t}$ is the 
consequence of the tensor meson
current conservation 
(\ref{transverse}). In the classical Regge scheme
\begin{equation}
\sum_J (2J+1){\cal M}_J {\cal P}_J(-z_t)\to
\eta_{\mathbb R}(t)
\beta_{\mathbb R}(t) \left( \frac{s}{s_0}\right)^{\alpha_{\mathbb R}(t)},
\label{eq:ReggeClassic}
\end{equation}
where this factor is absorbed into the unknown 
residue $\beta_{\mathbb R}(t)$. In our prescription t dependence
of the residue looks like
\begin{equation}
\label{eq:ourResidue}
\beta_{\mathbb R}\sim T_{01}(t)T_{02}(t) (-t)^{\alpha_{\mathbb R}/2}.
\end{equation}
There is no zero in t, since the Regge approach is valid only for
$|z_t|\gg 1$. But sometimes this behaviour at small t is extracted
in an explicit form like in the paper~\cite{DLSD} devoted to the process of
single diffraction dissociation.

\section*{Appendix B}

\begin{figure}[h!]  
 \includegraphics[width=0.49\textwidth]{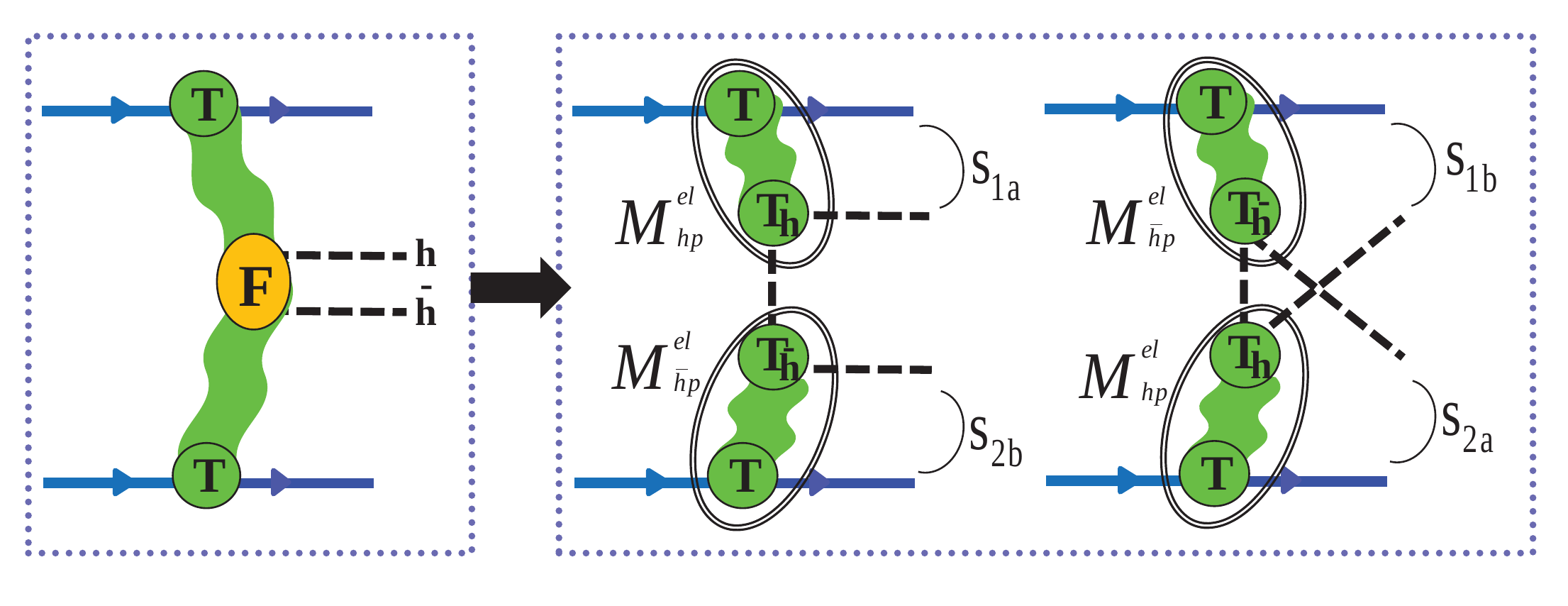} 
  \caption{\label{fig:MBHH} Scheme of calculation of the ``bare'' EDDE amplitude
  in the case of the low invariant mass ($M<3$~GeV) di-hadron production. Elastic 
  amplitudes are shown here as reggeon exchanges enclosed in ellipses.}
\end{figure}

Here we present general structure of the amplitude for
the $2\to 4$ process $p+p\to p+h\bar{h}+p$. This amplitude is
depicted in the Fig.~\ref{fig:MBHH}. Definitionss for the 
kinematics are
\begin{eqnarray}
s_{1\{a,b\}}&=&(p_{1}^{\prime}+k_{a,b})^2,\; s_{2\{a,b\}}=(p_{2}^{\prime}+k_{a,b})^2,\nonumber\\
\hat{t}_{a,b}&=&(p_1-p_1^{\prime}-k_{a,b})^2=(p_2-p_2^{\prime}-k_{b,a})^2,\label{kin:invarsab}
\end{eqnarray}

Leading contribution to the reggeon-reggeon fusion vertex is given by
the exchange amplitude (the right part of the Fig.~\ref{fig:MBHH}). In this case
we can write
\begin{eqnarray}
&& {\cal M}_{h\bar{h}}=
{\cal M}^{el}_{hp}(s_{1a},t_1)
\frac{\left(F_h(\hat{t}_a)\right)^2}{(\hat{t}_a-m_0^2)}
{\cal M}^{el}_{\bar{h}p}(s_{2b},t_2)+\nonumber\\
&& \label{eq:MBhh} 
{\cal M}^{el}_{hp}(s_{1b},t_1)
\frac{\left(F_h(\hat{t}_b)\right)^2}{(\hat{t}_b-m_0^2)}
{\cal M}^{el}_{\bar{h}p}(s_{2a},t_2).
\end{eqnarray}
Here ${\cal M}^{el}_{hp}$ and ${\cal M}^{el}_{\bar{h}p}$ are amplitudes of
the elastic hadron-proton scattering, which can be evaluated in any
appropriate approach, $F_h$ is the formfactor taking into account
the off-shellness of the exchanged hadron. For example, we can use 
simple reggeon exchanges 
for these amplitudes as was done 
in~\cite{EDDEthHH2a},\cite{EDDEthHH3}. Strictly 
speaking, we have to
take into account rescattering (unitarity) corrections 
since $s_{i\{a,b\}}$ can be of the order $\sim\sqrt{s}$. 

Calculations for the $\pi^+\pi^-$ production in this paper are
based on the simple regge formula (as in~\cite{EDDEthHH2a},\cite{EDDEthHH3}) 
for pion-proton elastic amplitudes
\begin{eqnarray}
&& \label{Melpip} {\cal M}^{el}_{\pi^{\pm}p}(s,t)=\sum_{i={\mathbb P},{\mathbb R}}
C^{\pm}_i F_i(t)\left( \frac{s}{s_0}\right)^{\alpha_i(t)},\\
&& \label{Formpip} F_i(t)={\mathrm e}^{B^0_i t},\; 
\alpha_i(t)=\alpha_i^0+\alpha_i^{\prime}t,
\end{eqnarray}
\begin{eqnarray}
&& C_{{\mathbb P}}=\mathrm{i} C_P,\; 
C^{\pm}_{{\mathbb R}}=C_f(a_f+\mathrm{i})\pm C_{\rho}(a_{\rho}-\mathrm{i}),\nonumber\\
&& C_P=13.63\;{\mathrm mb},\; C_f=31.79\;{\mathrm mb},\; 
C_{\rho}=4.23\;{\mathrm mb},\nonumber\\
&& a_f=-0.860895,\; a_{\rho}=-1.16158,\\
&& B^0_{{\mathbb P}}=2.75\;{\mathrm GeV}^{-2},\;
B^0_{{\mathbb R}}=2\;{\mathrm GeV}^{-2},\\
&& \label{Melpars} \alpha_{\mathbb P}(t)=1.088+0.25t,\; 
\alpha_{\mathbb R}=0.5475+0.93t,\\
&& F_{\pi}(\hat{t})\simeq 
{\mathrm e}^{\frac{\hat{t}-m_{\pi}^2}{\Lambda^2_{eff}}},\;
\Lambda^2_{eff}=1\;{\mathrm GeV}^2.
\end{eqnarray}

\section*{Appendix C}

Here we calculate the hadron-hadron ``soft'' interaction in the 
initial and in the final 
states (unitary corrections or re-scattering). It is denoted 
by $V$ in Fig.~\ref{fig1:lowmalg} and given 
by the following analytical
expressions:
\begin{eqnarray}
&&{\cal M}^{U}(p_1, p_2, \Delta_1, \Delta_2) = \int \frac{d^2\vec{q}_T}{(2\pi)^2} \,
\frac{d^2\vec{q}^{\;\prime}_T}{(2\pi)^2} \; V(s, \vec{q}_T) \;
\nonumber \\
&&\label{MUgen} \times {\cal M}( p_1-q_T, p_2+q_T,\Delta_{1T}, \Delta_{2T}) \,
V(s^{\prime}, \vec{q}^{\;\prime}_T) \;,
\\
&&\label{Vblobs} V(s, \vec{q}_T) = \int d^2\vec{b} \, {\mathrm e}^{i\vec{q}_T
\vec{b}} \sqrt{1+2\mathrm{i} T^{el}_{pp\to pp}(s, \vec{b})},
\end{eqnarray}
where $\Delta_{1T} = \Delta_{1} -q_T - q^{\prime}_T$, $\Delta_{2T}
= \Delta_{2} + q_T + q^{\prime}_T$, ${\cal M}$ is the ``bare'' amplitude of the
process $p+p\to p+M+p$. In the case of the eikonal representation of the
elastic amplitude $T^{el}_{pp\to pp}$ we have
\begin{equation}
\label{eq:Veik}
V(s, \vec{q}_T) = \int d^2\vec{b} \, \mathrm{e}^{\mathrm{i}\vec{q}_T
\vec{b}} \mathrm{e}^{\mathrm{i}\delta_{pp\to pp}(s,\vec{b})},
\end{equation}
where $\delta_{pp\to pp}$ is the eikonal function. In this case
amplitude~(\ref{MUgen}) can be rewritten as
\begin{eqnarray}
 {\cal M}^{U}(\vec{\Delta}_1, \vec{\Delta}_2) &=&
\int \frac{d^2\vec{b}}{2\pi} \, {\mathrm e}^{
-{\mathrm i}\vec{\delta}\vec{b}-\Omega(s,b)-\Omega(s^{\prime},b)
}\times\nonumber\\
&& 
\int \frac{d^2\vec{\kappa}}{2\pi} \, {\mathrm e}^{{\mathrm i}\vec{\kappa}\vec{b}}
{\cal M}(\vec{\Delta}-\vec{\kappa}, \vec{\Delta}+\vec{\kappa}),\nonumber\\
 \Omega(s,b)&=&-\mathrm{i}\delta_{pp\to pp}(s,b),\nonumber\\
 \vec{\Delta}&=&\frac{\vec{\Delta}_2+\vec{\Delta}_1}{2},\; 
\vec{\delta}=\frac{\vec{\Delta}_2-\vec{\Delta}_1}{2},\;\nonumber\\
\label{MUeik}\vec{\kappa}&=&\vec{\delta}+\vec{q}_T+\vec{q}^{\;\prime}_T.
\end{eqnarray}

Here we use the following representation for the elastic amplitude~\cite{3Pom}
\begin{equation}
\label{eq:3Prepresentation}
T_{el}(s,b)=\imath\left( 1-\mathrm{e}^{-2\Omega(s,b)}\right)/2.
\end{equation}
Some other groups~\cite{KMReik},\cite{GLMeik} use another conventions
\begin{equation}
T_{el}(s,b)=\imath\left( 1-\mathrm{e}^{-\Omega(s,b)/2}\right),
\end{equation}
or~\cite{BSWeik}
\begin{equation}
T_{el}(s,b)=\imath\left( 1-\mathrm{e}^{-\Omega(s,b)}\right),
\end{equation}
which are mathematically equivalent.

Let us consider calculations for concrete expressions of ${\cal M}$. To 
explore general features of diffractive patterns for the eikonal function 
we take the model~\cite{3P2O} (which originates from~\cite{3Pom} and uses
simple eikonal approximation) 
as an example. Nevertheless, some authors~\cite{KMReik},\cite{GLMeik2} point out 
that we have to use multichannel
eikonals to take into account multiple diffractive eigenstates. Here we 
have to point out that the parametrization~(\ref{eq:3Prepresentation}) 
satisfies exactly the unitarity condition 
and can be used without consideration of
any inner structure of the eikonal (diffractive eigenstates) 
as it was done in the
multichannel approach. We assume that the 
model~\cite{3P2O} is rather good for our purposes, at 
least for $|t_i|<1.5\;\mathrm{GeV}^2$, since it describes well
the latest data~\cite{lastTOTEM}.

For the amplitude we perform calculations
for several cases:
\begin{eqnarray}
 {\cal M}_i(\vec{\Delta}_1, \vec{\Delta}_2)&=&
 {\cal H}(\vec{\Delta}^2)
 {\mathrm e}^{-B_1\vec{\Delta}_1^2-B_2\vec{\Delta}_2^2}
{\cal K}_i(\vec{\Delta}_1, \vec{\Delta}_2)\to\nonumber\\
&& \label{Mvariants}\hspace*{-2.5cm}{\cal H}(\vec{\Delta}^2) 
{\mathrm e}^{
-B_+\left(\vec{\Delta}^2+\vec{\kappa}^2\right)+2B_-\left(\vec{\Delta}\vec{\kappa}\right)
}
{\cal K}_i(\vec{\Delta}-\vec{\kappa}, \vec{\Delta}+\vec{\kappa}),\\
 B_{\pm}&=&B_1\pm B_2,\nonumber\\
 \label{vK0}{\cal K}_0&=&1,\;\\
 \label{vKV}
{\cal K}_V&=&
\left[\vec{\Delta}_1\times\vec{\Delta}_2\right]\to 2\left[\vec{\Delta}\times\vec{\kappa}\right],\\
 \label{vKS}
{\cal K}_S&=&
\left(\vec{\Delta}_1\vec{\Delta}_2\right)\to \vec{\Delta}^2-\vec{\kappa}^2,\\
\label{vKT}
{\cal K}_T&=&
\vec{\Delta}_1^2\vec{\Delta}_2^2\to \left(\vec{\Delta}^2+\vec{\kappa}^2\right)^2-
4\left(\vec{\Delta}\vec{\kappa}\right)^2.
\end{eqnarray}
We have to calculate the following auxiliary integrals:
\begin{eqnarray}
 \label{Ikapv}{\cal I}_{\vec{\kappa}}^v&=&\int \frac{d^2\vec{\kappa}}{2\pi}\;  
{\mathrm e}^{
{\mathrm i}\vec{\kappa}\vec{b}-B_+\vec{\kappa}^2+2B_-\left(\vec{\Delta}\vec{\kappa}\right)
}\; v,\\
 {\cal I}_{\vec{\kappa}}^1&=&\frac{1}{2B_+}{\mathrm e}^{a/B_+},\nonumber \\
 \label{Ikap1}a&=&
 B_-^2\vec{\Delta}^2+{\mathrm i}B_-\left(\vec{\Delta}\vec{b}\right)-b^2/4,\\
 \label{Ikapkapi}{\cal I}_{\vec{\kappa}}^{\vec{\kappa}_{(i)}}&=&
-{\mathrm i}\frac{\partial}{\partial \vec{b}_{(i)}} {\cal I}_{\vec{\kappa}}^1=
\left( 
\frac{B_-}{B_+}\vec{\Delta}_{(i)}+{\mathrm i}\frac{\vec{b}_{(i)}}{2B_+}
\right){\cal I}_{\vec{\kappa}}^1,\\
 \label{Ikapkap2}
{\cal I}_{\vec{\kappa}}^{\vec{\kappa}^2}&=&
-\frac{\partial}{\partial B_+}{\cal I}_{\vec{\kappa}}^1=
\frac{1}{B_+}\left( 1+\frac{a}{B_+}\right)
{\cal I}_{\vec{\kappa}}^1,\\
 \label{Ikapkap4}
{\cal I}_{\vec{\kappa}}^{\vec{\kappa}^4}&=&
-\frac{\partial}{\partial B_+}{\cal I}_{\vec{\kappa}}^{\vec{\kappa}^2}=
\frac{1}{B_+^2}\left( 2+\frac{4a}{B_+}+\frac{a^2}{B_+^2}\right)
{\cal I}_{\vec{\kappa}}^1.
\end{eqnarray}
\begin{figure}[h!]  
 \includegraphics[width=0.49\textwidth]{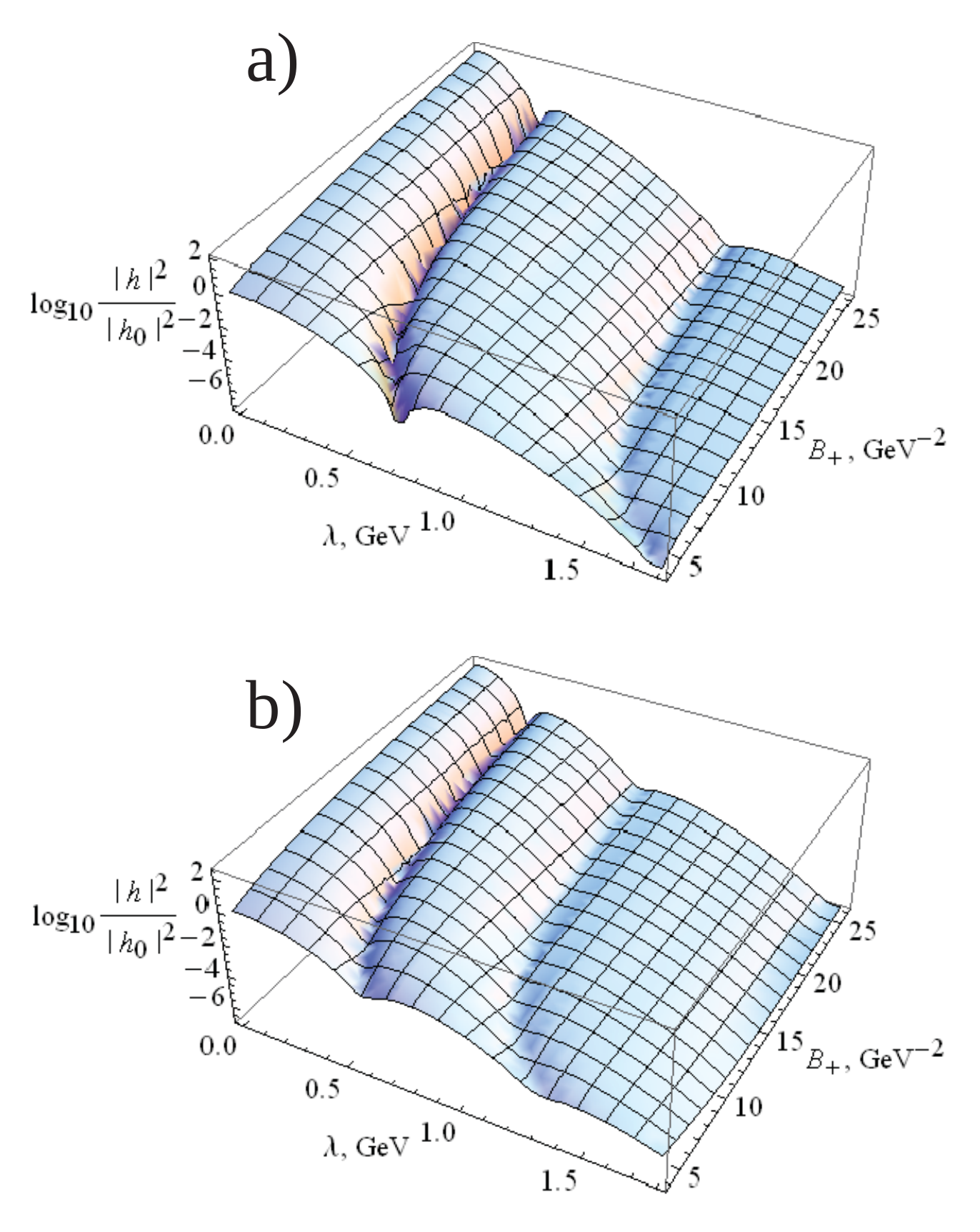} 
  \caption{\label{fig:hsquared} Function $|h(\lambda,B_+)|^2$ at a) $\sqrt{s}=62$~GeV and b) $\sqrt{s}=7$~TeV. $|h(0.,4.)|^2=20.5\;{\mathrm GeV}^{-2}$ at $\sqrt{s}=62$~GeV and $|h(0.,4.)|^2=5.27\;{\mathrm GeV}^{-2}$ at 
  $\sqrt{s}=7$~TeV.}
\end{figure}

Further calculations are expressed in terms of the function 
$h$ (see Fig.~\ref{fig:hsquared})
\begin{eqnarray}
&& h(\lambda,B_+)=\int\frac{d^2\vec{b}}{2\pi}
{\mathrm e}^{-\mathrm{i}\vec{\lambda}\vec{b}-\Omega(s,b)-\Omega(s^{\prime},b)-b^2/(4B_+)}=\nonumber\\
&& \phantom{h(\lambda,B_+)=}\int db\; b\; J_0(\lambda b) 
{\mathrm e}^{-\Omega(s,b)-\Omega(s^{\prime},b)-b^2/(4B_+)},\nonumber\\
&& \label{hUnitar}
\phantom{h(\lambda,B_+)}\lambda=\left|\vec{\lambda}\right|=\left|\vec{\delta}-\frac{B_-}{B_+}\vec{\Delta}\right|.
\end{eqnarray}

We can write
\begin{eqnarray}
 \label{MUvariants}
{\cal M}^U_i&=&\frac{1}{2B_+}{\cal H}(\vec{\Delta}^2)
{\mathrm e}^{-B_+\left(1-\frac{B_-^2}{B_+^2}\right)\vec{\Delta}^2}\hat{{\cal K}}_i h(\lambda,B_+),\\
 \label{vKU0}\hat{{\cal K}}_0&=&1,\\
 \label{vKUV}\hat{{\cal K}}_V&=&
-\frac{1}{B_+}\left[\vec{\Delta}\times\vec{\lambda}\right]
\frac{1}{\lambda}\frac{\partial}{\partial\lambda},\\
 \hat{{\cal K}}_S&=&
-\frac{1}{B_+}+
\vec{\Delta}^2\left(1-\frac{B_-^2}{B_+^2}\right)+\nonumber\\
 \label{vKUS} 
 &\phantom{+}& \frac{B_-}{B_+^2}\left( \vec{\Delta}\vec{\lambda}\right)
 \frac{1}{\lambda}\frac{\partial}{\partial\lambda}+ 
\frac{\partial}{\partial B_+},\\
 \hat{{\cal K}}_T&=&
\left[
\frac{2}{B_+^2}+\frac{2\vec{\Delta}^2}{B_+}\left( 1+\frac{2B_-^2}{B_+^2}\right)+
\vec{\Delta}^4\left(1-\frac{B_-^2}{B_+^2}\right)
\right]+\nonumber\\
&\phantom{+}& \left[
\frac{2B_-}{B_+^2}\left( -\frac{2}{B_+}+\vec{\Delta}^2\left( 1-\frac{B_-^2}{B_+^2}\right)\right)
\left( \vec{\Delta}\vec{\lambda}\right)-\right.\nonumber\\
&\phantom{-}&\left.
\frac{\vec{\Delta}^2}{B_+^2}\left( 1-\frac{B_-^2}{B_+^2}\right)
\right]\frac{1}{\lambda}\frac{\partial}{\partial\lambda}-
\nonumber\\
&\phantom{-}&
2\left( 
\frac{1}{B_+}+\vec{\Delta}^2\left( 1+\frac{B_-^2}{B_+^2}\right)
\right)\frac{\partial}{\partial B_+}+
\nonumber\\
&\phantom{+}&
\frac{\partial^2}{\partial^2 B_+}+
2\frac{B_-}{B_+^2}\left( \vec{\Delta}\vec{\lambda}\right)
\frac{1}{\lambda}\frac{\partial}{\partial\lambda}
\frac{\partial}{\partial B_+}-\nonumber\\
&\phantom{-}&
\frac{1}{B_+^2}\left( 1-\frac{B_-^2}{B_+^2}\right)
\left(\vec{\Delta}\vec{\lambda}\right)^2
\frac{1}{\lambda}\frac{\partial}{\partial\lambda}
\frac{1}{\lambda}\frac{\partial}{\partial\lambda}
\label{vKUT}.
\end{eqnarray}

The ratio
\begin{equation}
\label{eq:SoftSurv0}
<S^2>=\frac{\int\int d^2\vec{\Delta}_1 d^2\vec{\Delta}_2 
\left| {\cal M}^{U}\right|^2}{\int\int d^2\vec{\Delta}_1 d^2\vec{\Delta}_2 \left| {\cal M}\right|^2}
\end{equation}
is usually called ``soft survival probability''. For example, at 
$\sqrt{s}=14$~TeV the value of $<S^2>$ is about $0.03$ for the slope
of t-distribution $\sim 4\;{\mathrm GeV}^{-2}$ 
(invariant masses about $100$~GeV) and
$0.13$ for the slope $\sim 10\;{\mathrm GeV}^{-2}$ 
(invariant masses about $1$~GeV).

\section*{Aknowledgements}

Author thanks to V.~A.~Petrov and A.~A.~Godizov for useful discussions.

%

\begin{thebibliography}{}
%
%

\bibitem{EDDEthOLD1} T.W.~Kibble, Proc. Roy. Soc. \textbf{244}, 355 (1958)
\bibitem{EDDEthOLD2} A.A.~Logunov, A.N.~Tavkhelidze, Nucl. Phys. \textbf{8}, 374 (1958)  
\bibitem{EDDEthOLD3} S.S.~Gershtein, A.A.~Logunov, 
\emph{Growth Of Hadron Cross-sections And Its Possible Connection With Glueballs},
Sov. J. Nucl. Phys. \textbf{39}, 960 (1984) [Yad. Fiz. \textbf{39}, 1514 (1984)]
\bibitem{EDDEthOLD4} K.A.~Ter-Martirosyan, Nucl. Phys. \textbf{68}, 591 (1964)  
\bibitem{EDDEthOLD5} K.G.~Boreskov, Yad. Fiz. \textbf{8}, 796 (1968) 
\bibitem{EDDEthOLD6} A.~Actor,
\emph{Characteristics of double-pomeron exchange},
Ann. of Phys. \textbf{109}, 317 (1977)  
\bibitem{EDDEthOLD7} J.~Pumplin, F.S.~Henyey,
\emph{Double pomeron exchange in the reaction $pp \to pp\pi^+\pi^-$},
Nucl. Phys. B \textbf{117}, 377 (1976)  
\bibitem{EDDEthOLD8} A.~Bialas, P.V.~Landshoff,
\emph{Higgs production in p p collisions by double pomeron exchange}
Phys. Lett. B \textbf{256}, 540 (1991)  
\bibitem{EDDEthOLD9} B.R.~Desai, B.C.~Shen, M.~Jacob, 
\emph{Double pomeron exchange in high-energy pp collisions},
Nucl. Phys. B \textbf{142}, 258 (1978)    
\bibitem {EDDEexpOLD1} WA102 Collaboration, 
\emph{A Study of pseudoscalar states produced centrally in p p interactions at 450~GeV/c},
Phys. Lett. B \textbf{427}, 398 (1998) 
\bibitem {EDDEexpOLD2} WA102 Collaboration,
\emph{Experimental evidence for a vector like behavior of Pomeron exchange},
Phys. Lett. B \textbf{467}, 165 (1999)  
\bibitem {EDDEexpOLD3} WA102 Collaboration,
\emph{A Study of the f(0)(1370), f(0)(1500), f(0)(2000) and f(2)(1950) observed in the centrally produced 4$\pi$ final states},
Phys. Lett. B \textbf{474}, 423 (2000)  
\bibitem {EDDEexpOLD4} WA102 Collaboration,
\emph{A Coupled channel analysis of the centrally produced K+ K- and pi+ pi- final states in p p interactions at 450~GeV/c},
Phys. Lett. B \textbf{462}, 462 (1999)  
\bibitem {EDDEexpOLD5} A.~Kirk,
\emph{Resonance production in central p p collisions at the CERN Omega spectrometer},
Phys. Lett. B \textbf{489}, 29 (2000)
\bibitem{EDDEexpOLD6} A.~Breakstone {\it et al.}  [Ames-Bologna-CERN-Dortmund-Heidelberg-Warsaw Collaboration],
\emph{The Reaction pomeron-pomeron ---> pi+ pi- and an unusual production mechanism for the f2 (1270)},
Z. Phys. C \textbf{48}, 569 (1990)
\bibitem{KMR2014} L.A.~Harland-Lang, V.A.~Khoze, M.G.~Ryskin and W.J.~Stirling,
\emph{Central exclusive production within the Durham model: a review},
Int. J. Mod. Phys. A \textbf{29}, 1430031 (2014)
\bibitem{my2013} R.A.~Ryutin,
\emph{Exclusive double diffractive events: general framework and prospects},
Eur. Phys. J. C \textbf{73}, 2443 (2013)
\bibitem{EDDEth1} L.~A.~Harland-Lang, V.~A.~Khoze, M.~G.~Ryskin and W.~J.~Stirling,
\emph{Latest Results in Central Exclusive Production: A Summary},
arXiv:1301.2552 [hep-ph]
\bibitem{EDDEth2} L.A.~Harland-Lang, V.A.~Khoze, M.G.~Ryskin, W.J.~Stirling, 
\emph{The Phenomenology of Central Exclusive Production at Hadron Colliders},
 Eur. Phys. J. C \textbf{72}, 2110 (2012) 
\bibitem{EDDEth3} V.A.~Khoze, A.D.~Martin, M.G.~Ryskin,	
\emph{New Physics with Tagged Forward Protons at the LHC},
Frascati Phys. Ser. \textbf{44}, 147 (2007) 
\bibitem{EDDEth4} L.A.~Harland-Lang, V.A.~Khoze, M.G.~Ryskin, W.J.~Stirling 	
\emph{Standard candle central exclusive processes at the Tevatron and LHC},
Eur. Phys. J. C \textbf{69}, 179 (2010) 
\bibitem{EDDEth5} V.A.~Petrov, R.A.~Ryutin,
\emph{Exclusive double diffractive events: Menu for LHC}, 
JHEP \textbf{0408}, 013 (2004) 
\bibitem{EDDEth6} V.A.~Petrov, R.A.~Ryutin, 	
\emph{Patterns of the exclusive double diffraction},
J. Phys. G \textbf{35}, 065004 (2008) 
\bibitem{EDDEth7} M.G.~Albrow, T.D.~Coughlin, J.R.~Forshaw,  	
\emph{Central Exclusive Particle Production at High Energy Hadron Colliders},
Prog. Part. Nucl. Phys. \textbf{65}, 149 (2010)
\bibitem{EDDEth8} R.~Enberg, G.~Ingelman, N.~Timneanu,
\emph{Soft color interactions and diffractive Higgs production},
Eur. Phys. J. C \textbf{33}, S542 (2004)
\bibitem{EDDEth9} E.~Gotsman, H.~Kowalski, E.~Levin, U.~Maor, A.~Prygarin,
\emph{Survival probability for diffractive dijet production at the LHC},
Eur. Phys. J. C \textbf{47}, 655 (2006)  
\bibitem{EDDEth10} S.M.~Troshin, N.E.~Tyurin,
\emph{Reflective scattering effects in double-pomeron exchange processes},
Mod. Phys. Lett. A \textbf{23}, 169 (2008)   
\bibitem{EDDEth11} C.P.~Herzog, S.~Paik, M.J.~Strassler, E.G.~Thompson,
\emph{Holographic Double Diffractive Scattering},
JHEP \textbf{0808}, 010 (2008)
\bibitem{MT2014} M.~Tasevsky, 
\emph{Review of Central Exclusive Production of the Higgs Boson Beyond the Standard Mode}, 
arXiv:1407.8332 [hep-ph], to be published in Int. J. Mod. Phys. A
\bibitem{EDDEthH1} S.~Heinemeyer, V.~A.~Khoze, M.~G.~Ryskin, W.~J.~Stirling, M.~Tasevsky and G.~Weiglein,
\emph{Central Exclusive Diffractive MSSM Higgs-Boson Production at the LHC},
J. Phys. Conf. Ser. \textbf{110}, 072016 (2008)
\bibitem{EDDEthH2} M.~Chaichian, P.~Hoyer, K.~Huitu, V.~A.~Khoze and A.~D.~Pilkington,
\emph{Searching for the triplet Higgs sector via central exclusive production at the LHC},
JHEP \textbf{0905}, 011 (2009)
\bibitem{EDDEthH3} J.R.~Cudell, A.~Dechambre, O.F.~Hernandez,	
\emph{Higgs Central Exclusive Production},
Phys. Lett. B \textbf{706}, 333 (2012)
\bibitem{EDDEthH4} V.A.~Petrov, R.A.~Ryutin,
\emph{Exclusive double diffractive Higgs boson production at LHC}, 
Eur. Phys. J. C\textbf{36}, 509 (2004)
\bibitem{EDDEthH5} M.~B.~G.~Ducati, M.~M.~Machado and G.~G.~Silveira,
\emph{Single and Central Diffractive Higgs Production at the LHC},
AIP Conf. Proc.  \textbf{1350}, 128 (2011)
\bibitem{EDDEthH6} R.~Enberg and R.~Pasechnik,
\emph{Associated central exclusive production of charged Higgs bosons},
Phys. Rev. D \textbf{83}, 095020 (2011)
\bibitem{EDDEthH7} E.~Levin and J.~Miller,
\emph{Central exclusive diffractive Higgs boson production in hadron-nucleus and nucleus-nucleus collisions at the LHC},
arXiv:0801.3593 [hep-ph]
\bibitem{EDDEthH8} R.C.~Brower, M.~Djuric, C.-I~Tan, 
\emph{Diffractive Higgs Production by AdS Pomeron Fusion}, 
JHEP \textbf{1209}, 097 (2012)
\bibitem{EDDEthH9} B.Z.~Kopeliovich , I.~Schmidt,    
\emph{Higgs diffractive production},
Nucl. Phys. A \textbf{782}, 118 (2007)
\bibitem{EDDEthH10} A.~Bzdak,	
\emph{Exclusive Higgs and dijet production by double pomeron exchange: The CDF upper limits},
Phys. Lett. B \textbf{615}, 240 (2005)
\bibitem{EDDEthH11} D.~Kharzeev, E.~Levin, 
\emph{Soft double diffractive Higgs production at hadron colliders}, 
Phys. Rev. D \textbf{63}, 073004 (2001)
\bibitem{Cox:2005if} B.E.~Cox, A.~De~Roeck, V.A.~Khoze, T.~Pierzchala, M.G.~Ryskin, I.~Nasteva, W.J.~Stirling and M.~Tasevsky,
\emph{Detecting the standard model Higgs boson in the WW decay channel using forward proton tagging at the LHC},
Eur. Phys. J. C \textbf{45}, 401 (2006)
\bibitem{Tasevsky:2013iea} M.~Tasevsky,
\emph{Exclusive MSSM Higgs production at the LHC after Run I},
Eur. Phys. J. C \textbf{73}, 2672 (2013)
\bibitem{Heinemeyer:2010gs} S.~Heinemeyer, V.A.~Khoze, M.G.~Ryskin, M.~Tasevsky and G.~Weiglein,
\emph{BSM Higgs Physics in the Exclusive Forward Proton Mode at the LHC},
Eur. Phys. J. C \textbf{71}, 1649 (2011)  
\bibitem{EDDEthQ1}  V.A.~Khoze, A.D.~Martin, M.G.~Ryskin, W.J.~Stirling, 
\emph{Double diffractive chi meson production at the hadron colliders}, 
Eur. Phys. J. C \textbf{35}, 211 (2004)
\bibitem{EDDEthHH1} L.A.~Harland-Lang, V.A.~Khoze, M.G.~Ryskin, W.J.~Stirling,
\emph{Central exclusive meson pair production in the perturbative regime at hadron colliders},
Eur. Phys. J. C \textbf{71}, 1714 (2011)
\bibitem{EDDEthHH2} L.~A.~Harland-Lang, V.~A.~Khoze, M.~G.~Ryskin and W.~J.~Stirling,
\emph{Central exclusive production as a probe of the gluonic component of the eta' and eta mesons},
Eur. Phys. J. C \textbf{73}, 2429 (2013)
\bibitem{EDDEthHH2a} L.~A.~Harland-Lang, V.~A.~Khoze, M.~G.~Ryskin,
\emph{Modeling exclusive meson pair production at hadron colliders},
arXiv:1312.4553 [hep-ph]
\bibitem{EDDEthHH3} R.~Staszewski, P.~Lebiedowicz, M.~Trzebinski, J.~Chwastowski and A.~Szczurek,
\emph{Exclusive $\pi^+\pi^-$ Production at the LHC with Forward Proton Tagging},
Acta Phys. Polon. B \textbf{42}, 1861 (2011)
\bibitem{EDDEexp1} M.~Albrow {\it et al.}  [CDF Collaboration],
\emph{Exclusive Central pi+pi- production in CDF},
arXiv:1310.3839 [hep-ex]
\bibitem{EDDEexp2} M.~Albrow,
\emph{Summary of the EDS Blois 2013 Workshop},
arXiv:1310.7047 [hep-ex]
\bibitem{EDDEexp3} F.~Reidt [ALICE Collaboration],
\emph{Central Diffraction in Proton-Proton Collisions at $\sqrt{s}=7$\,TeV with ALICE at LHC},
AIP Conf. Proc. \textbf{1523}, 17 (2012); arXiv:1301.3507 [hep-ex]
\bibitem{EDDEexp4} D.~Moran,
\emph{Central Exclusive Production with Dimuon Final States at LHCb},
CERN-THESIS-2011-209 (2011) 
\bibitem{EDDEexp5} G.A.~Alves et al. (for the CMS Collaboration), 
\emph{Search for central exclusive gamma pair production and observation of central exclusive electron pair production in pp collisions at $\sqrt{s}=7$~TeV},
CMS-PAS-FWD-11-004, CERN-PH-EP-2012-246 (2012), accepted for publication in JHEP. 
\bibitem{EDDEexp6} K.~Goulianos (CDF II Collaboration),
\emph{Diffraction Results from CDF},
arXiv:1204.5241 [hep-ex]
\bibitem{EDDEexp7} T.~Aaltonen et al. (CDF Collaboration),
\emph{Observation of Exclusive Dijet Production at the Fermilab Tevatron $p^- \bar{p}$ Collider},
Phys. Rev. D \textbf{77}, 052004 (2008) 
\bibitem{EDDEexp8} T.~Aaltonen et al. (CDF Collaboration),
\emph{Observation of Exclusive Gamma Gamma Production in $p \bar{p}$ Collisions at $\sqrt{s}=1.96$ TeV},
Phys. Rev. Lett. \textbf{108}, 081801 (2012) 
\bibitem{EDDEexp9} T.~Aaltonen et al. (CDF Collaboration),
\emph{Search for exclusive $\gamma \gamma$ production in hadron-hadron collisions},
Phys. Rev. Lett. \textbf{99}, 242002 (2007)
\bibitem{LRGs} J.D.~Bjorken, 
\emph{Rapidity gaps and jets as a new-physics signature in very-high-energy hadron-hadron collisions},
Phys. Rev. D \textbf{47}, 101 (1993)
\bibitem{LRGs2} F.~Abe et al. (CDF Collaboration),
\emph{Observation of rapidity gaps in $\bar{p}p$ collisions at 1.8~TeV},
Phys. Rev. Lett. \textbf{74}, 855 (1995)
\bibitem{MMM} M.G.~Albrow , A.~Rostovtsev,
\emph{Searching for the Higgs at hadron colliders using the missing mass method},
FERMILAB-PUB-00-173 (2000), arXiv: hep-ph/0009336 [hep-ph]
\bibitem{spin_parity_analyser4} V.A.~Petrov, R.A.~Ryutin,  A.E.~Sobol, J.-P.~Guillaud, 
\emph{Azimuthal angular distributions in EDDE as spin-parity analyser and glueball filter for LHC},
JHEP \textbf{0506}, 007 (2005) 
\bibitem{spin_parity_analyser2a} A.B.~Kaidalov, V.A.~Khoze, A.D.~Martin, M.G.~Ryskin, 
\emph{Central exclusive diffractive production as a spin-parity analyser: From Hadrons to Higgs},
Eur. Phys. J. C \textbf{31}, 387 (2003)
\bibitem{HERAevmp3} M.~Derrick et al. (ZEUS Collaboration), 
\emph{Measurement of elastic omega photoproduction at HERA}, 
Z. Phys. C \textbf{73}, 73 (1996)
\bibitem{HERAevmp3a} J.~Breitweg et al. (ZEUS Collaboration),	
\emph{Elastic and proton dissociative $\rho_0$ photoproduction at HERA}, 
Eur. Phys. J. C \textbf{2}, 247 (1998)
\bibitem{HERAevmp4} M.~Derrick et al. (ZEUS Collaboration),  
\emph{Measurement of elastic $\phi$ photoproduction at HERA}, 
Phys. Lett. B \textbf{377}, 259 (1996)
\bibitem{Morrow} R.A.~Morrow,
\emph{Construction of multi-Regge amplitudes by the Van Hove-Durand method},
Phys. Rev. D \textbf{18}, 2672 (1978)
\bibitem{spin_parity_analyser3a} F.E.~Close, G.A.~Schuller,
\emph{Central production of mesons: Exotic states versus pomeron structure},
Phys. Lett. B \textbf{458}, 127 (1999)
\bibitem{spin_parity_analyser3b} F.E.~Close, G.A.~Schuller, 
\emph{Evidence that the pomeron transforms as a nonconserved vector current},
Phys. Lett. B \textbf{464}, 279 (1999) 
\bibitem{spin_parity_analyser2b} V.A.~Khoze, A.D.~Martin, M.G.~Ryskin,
\emph{Physics with tagged forward protons at the LHC},
Eur. Phys. J. C \textbf{24}, 581 (2002)
\bibitem{instanton1} E.V.~Shuryak, I.~Zahed, 
\emph{Semiclassical double pomeron production of glueballs and eta-prime}, 
Phys. Rev. D \textbf{68}, 034001 (2003)
\bibitem{instanton3} J.~Ellis, D.~Kharzeev, 
\emph{The Glueball filter in central production and broken scale invariance}, 
Preprint CERN-TH-98-349, arXiv: hep-ph/9811222. 
\bibitem{instanton4} N.I.~Kochelev, 
\emph{Unusual properties of the central production of glueballs and instantons}, 
arXiv: hep-ph/9902203.    
\bibitem{lowmassmesonsLHC} M.V.T.~Machado,  	
\emph{Investigating the central diffractive f0(980) and f2(1270) meson production at the LHC},
Phys. Rev. D \textbf{86}, 014029 (2012)
\bibitem{diff2} V.~A.~Petrov,A.~V.~Prokudin,S.~M.~Troshin and N.~E.~Tyurin,
\emph{Novel features of diffraction at the LHC},
J.Phys. G {\bf 27}, 2225 (2001) 
\bibitem{Godizov2012} A.A.~Godizov,
\emph{Models of elastic diffractive scattering to falsify at the LHC},
PoS \textbf{IHEP-LHC-2011}, 005 (2012); arXiv:1203.6013 [hep-ph]
\bibitem{CMSTOTEM} CMS Collaboration,  
\emph{CMS-TOTEM event display: high-pT jets with two leading protons},
CMS-DP-2013-004 ; CERN-CMS-DP-2013-004.
\bibitem{CMSTOTEM1} CMS Collaboration (R.A.~Ciesielski),
\emph{Measurements of diffraction in p-p collisions in CMS},
CMS-CR-2013-207, talk presented on
21st International Workshop on Deep-Inelastic Scattering and Related Subjects, 
Marseilles, Provence, France, 22 - 26 Apr 2013.
\bibitem{CMSTOTEM2} S.~Bertolucci, T.~Camporesi, S.~Giani,
\emph{CMS-TOTEM Join Project MoU},
CERN-RRB-2014-002.
\bibitem{add1} M.G.~Albrow,
\emph{High precision spectrometers for very forward protons in CMS},
AIP Conf. Proc. \textbf{1523}, 320 (2012).  
\bibitem{add2} M.~Tasevsky,
\emph{Diffractive physics program in ATLAS experiment},
Nucl. Phys. Proc. Suppl.  \textbf{179-180}, 187 (2008)
\bibitem{add3} M.~Tasevsky [ATLAS Collaboration],
\emph{Diffraction and central exclusive production at ATLAS},
AIP Conf. Proc. \textbf{1350}, 164 (2011).
\bibitem{lastTOTEM} G.~Antchev \emph{et al.}  [TOTEM Collaboration],
 \emph{Measurement of proton-proton elastic scattering and total cross-section at S**(1/2) = 7-TeV},
Europhys. Lett. \textbf{101}, 21002 (2013).
\bibitem{DLSD} A.~Donnachie and P.V.~Landshoff,
\emph{Soft diffraction dissociation}, 
ArXiv: hep-ph/0305246.
\bibitem{3Pom} V.A.~Petrov, A.V.~Prokudin, 
\emph{The First three pomerons....},
Eur. Phys. J. C \textbf{23}, 135 (2002)
\bibitem{KMReik} V.A.~Khoze, A.D.~Martin and M.G.~Ryskin,
 \emph{High Energy Elastic and Diffractive Cross Sections},
  Eur. Phys. J. C \textbf{74}, 2756 (2014)
\bibitem{GLMeik} E.~Gotsman, E.~Levin and U.~Maor,
\emph{Proton-air collisions in a model of soft interactions at high energies},
Phys. Rev. D \textbf{88}, 114027 (2013)
\bibitem{BSWeik} C.~Bourrely,
\emph{An analysis of elastic scattering reactions with a Fermi-Dirac pomeron opaqueness in impact parameter space},
Eur. Phys. J. C \textbf{74}, 2736 (2014)
\bibitem{3P2O} A.~Alkin, O.~Kovalenko and E.~Martynov,
\emph{Can the "standard" unitarized Regge models describe the TOTEM data?},
Europhys. Lett. \textbf{102}, 31001 (2013)
\bibitem{GLMeik2} E.~Gotsman, E.~Levin and U.~Maor,
\emph{Description of LHC data in a soft interaction model},
  Phys. Lett. B \textbf{716}, 425 (2012)
  
\end{thebibliography}
%

\end{document}